\documentclass[sigplan,nonacm]{acmart}
\usepackage{graphicx}
\usepackage{subcaption}
\usepackage{setspace} 
\usepackage{multirow}
\usepackage{caption}    
\usepackage{tikz}
\usepackage{xcolor}

\newcommand*\colorcircled[2]{%
  \tikz[baseline=(char.base)]{
    \node[shape=circle, fill=#1, minimum size=11pt, inner sep=0pt] (char)
      {\resizebox{!}{5pt}{\textcolor{black}{#2}}};}}

\newcommand*\greencircled[1]{\colorcircled{green!70!black}{#1}}
\newcommand*\bluecircled[1]{\colorcircled{cyan!80}{#1}}
\newcommand*\orangecircled[1]{\colorcircled{yellow!55!orange}{#1}}

\AtBeginDocument{%
  }

\makeatletter
\def\@copyrightspace{\relax}
\makeatother
\begin{document}

\title{Taiji: A DPU Memory Elasticity Solution for In-production Cloud Environments}

\author{Hao Zheng\textsuperscript{1*},\quad
Longxiang Wang\textsuperscript{2*},\quad
Yun Xu\textsuperscript{1},\quad
Qiang Wang\textsuperscript{2},\quad
Yibin Shen\textsuperscript{1},\quad
Xiaoshe Dong\textsuperscript{2},\quad
Bang Di\textsuperscript{1},\quad
Jia Wei\textsuperscript{2},\quad
Shenyu Dong\textsuperscript{1},\quad
Xingjun Zhang\textsuperscript{2},\quad
Weichen Chen\textsuperscript{1},\quad
Zhao Han\textsuperscript{3},\quad
Sanqian Zhao\textsuperscript{1},\quad
Dongdong Huang\textsuperscript{1},\quad
Jie Qi\textsuperscript{1},\quad
Yifan Yang\textsuperscript{1},\quad
Zhao Gao\textsuperscript{1},\quad
Yi Wang\textsuperscript{1},\quad
Jinhu Li\textsuperscript{1},\quad
Xudong Ren\textsuperscript{1},\quad
Min He\textsuperscript{1},\quad
Hang Yang\textsuperscript{1},\quad
Xiao Zheng\textsuperscript{1},\quad
Haijiao Hao\textsuperscript{1},\quad
Jiesheng Wu\textsuperscript{1}}
\thanks{\textsuperscript{*} Hao Zheng and Longxiang Wang contributed equally to this work.}

\affiliation{
\textsuperscript{1} Alibaba Cloud, Hangzhou, \country{China} \\
\textsuperscript{2} School of Computer Science, Xi'an Jiaotong University, Xi'an, \country{China} \\
\textsuperscript{3} Department of Computer Science, Shanghai Jiao Tong University, Shanghai, \country{China}
}




\begin{abstract}
The growth of cloud computing drives data centers toward higher density and efficiency. Data processing units (DPUs) enhance server network and storage performance but face challenges such as long hardware upgrade cycles and limited resources. To address these, we propose Taiji, a resource-elasticity architecture for DPUs. Combining hybrid virtualization with parallel memory swapping, Taiji switches the DPU's operating system (OS) into a guest OS and inserts a lightweight virtualization layer, making nearly all DPU memory swappable. It achieves memory overcommitment for the switched guest OS via high-performance memory elasticity, fully transparent to upper-layer applications, and supports hot-switch and hot-upgrade to meet in-production cloud requirements. Experiments show that Taiji expands DPU memory resources by over 50\%, maintains virtualization overhead around 5\%, and ensures 90\% of swap-ins complete within 10\textmu s. Taiji delivers an efficient, reliable, low-overhead elasticity solution for DPUs and is deployed in large-scale production systems across more than 30,000 servers.
\end{abstract}




\maketitle
\pagestyle{plain}

\section{Introduction}

To achieve low latency and high throughput in I/O processing, major cloud providers deploy DPUs to offload network, storage, and control tasks, allowing CPUs to focus on computation and improving system performance and efficiency \cite{RN35, 288657}. However, rapid growth in e-Commerce, mobile and AI applications drives an exponential demand for I/O, memory, and computing resources \cite{RN36}, increasing the requirements of DPU resources. Rising VM demands for peak IOPS and bandwidth, combined with higher VM density on cloud servers, strain DPU resource allocation. Current mainstream DPU models provide only 32GB of memory (e.g., BlueField-3 \cite{BlueFieldDPU_Blog}), struggling to meet growing service demands. Furthermore, DPU hardware upgrades are impractical due to long cycles and high costs in deployed systems \cite{nickel2024survey}.

Our observations in the cloud show that compute nodes rarely fully use their maximum storage IOPS and network PPS capabilities, with simultaneous peak utilization rare. DPU resources reserved for peak demand are often underutilized \cite{shah2023dag,shah2021dpu,shu2024burstable}. Memory is the primary infrastructure cost in current data centers \cite{weiner2022tmo}. Our production data indicates that a page accessed in 10 minutes is hot, with more than 90\% of DPUs having a cold page water level above 50\%. Thus, resource elasticity is a viable approach to address the mismatch between DPU hardware and software resource demands. Memory elasticity can be achieved through swapping or virtualization, but existing solutions face challenges in DPU:

1) \textbf{Memory swapping functionality and efficiency need enhancement.} High-performance DPU services, such as storage and networking, heavily rely on HugeTLB huge pages. Current mainstream solutions mainly support small pages and have limited large-scale deployment \cite{jo2023toward, song2016efficient,zhang2017memflex,tabatabai2024fbmm,sha2021swift}. Newer kernels support THP for huge page swapping, but splitting THP during swap-out impedes later huge page allocation \cite{arxiv_flexible_swapping_lwn,ren2024mtm,panwar2018making}. Moreover, a significant portion of kernel memory remains unswappable \cite{kernel_hugetlb_noswap}.

2) \textbf{High resource overhead and adaptation cost of virtualization solutions.} Virtualization enables memory swapping in VMs but incurs substantial resource costs beyond performance impact \cite{pearce2013virtualization,zhang2020giantvm,barham2003xen,velte2009microsoft,haletky2011vmware}. Both Type 1 hypervisors requiring a privileged VM and Type 2 hypervisors relying on a Host OS must run an additional OS, which is costly under DPU resource constraints \cite{virtualization_overhead_icpp, vmware_swap_file}. Migrating existing DPU workloads into a virtualized environment further requires high adaptation effort and risks lengthening I/O paths, potentially degrading native performance \cite{virtualization_overhead_icpp}.

Therefore, a lightweight software solution is needed to address DPU resource insufficiency. We present Taiji, a lightweight system for DPU resource elasticity. Taiji inserts a minimal virtualization layer between the host OS and hardware at runtime, enabling full hardware control while delivering efficient elastic resources to the OS. The design ensures no performance degradation, meets growing resource demands, and supports seamless deployment across large-scale, existing DPU servers. To our knowledge, this is the first paper to use virtualization for memory elasticity in DPUs. Our key contributions are:

\textbf{Lightweight Hybrid Virtualization on DPUs.} We propose a lightweight hybrid virtualization architecture for DPUs that achieves full memory swappability with minimal virtualization and resource overhead. By specialized address-space management, it switches the Host OS to a Guest OS and inserts a low-overhead virtualization layer without increasing resource overhead on the Host OS. It also minimizes resource overhead by sharing the user-space layer with the Guest OS, avoiding an extra user-space layer.

\textbf{High Performance DPU Memory Elasticity.} Targeting the widely used huge page in DPU environments, a high performance memory elasticity solution is designed at the huge page granularity, expanding virtual memory by more 50\%. Furthermore, considering the characteristics of the IO path in DPUs, we dynamically filter the potentially involved memory in DMA and provide efficient memory elasticity through parallelized LRU and SWAP design, ensuring rapid swap-in, with 90\% of page faults resolved in under 10 \textmu s.

\textbf{Seamless DPU resource elasticity scheduling.} Leveraging the distinct roles of data-plane processors (DPs) and control-plane processors (CPs) on DPUs, Taiji uses the CPs to allocate compute resources to elasticity tasks via priority coordination and fair time-slice allocation. This schedules front-end CPU and back-end memory elasticity tasks while minimizing interference with DP high-performance I/O. The unified scheduler also reserves interfaces for extending CPU resource elasticity.

\textbf{Large-scale Online DPU Deployment Practice.} To ensure compatibility with existing running DPUs, Taiji uses a modular design that needs no changes to the running OS, allowing seamless hot-switch of running DPUs while maintaining normal service operation and enabling elastic resource capabilities. Its hot-upgrade feature also enables rapid deployment, updates, and evolution, enhancing the reliability and scalability of DPU resource elasticity. Taiji is currently running stably online, deployed at scale on over 30,000 DPUs.

\section{Background and Motivation}

\subsection{Current Resource Usage of DPUs}
\subsubsection{DPU Usage Characteristics}
\begin{figure}[h!]
    \centering
    \includegraphics[width=\linewidth]{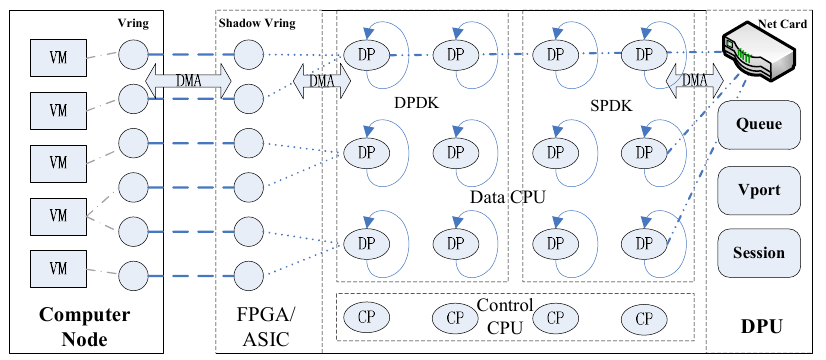}
    \caption{The DPU Resource Usage}
    \label{fig:dpu}
\end{figure}

In cloud computing, the \texttt{virtio} protocol \cite{zhang2024hd} is widely used in the I/O path, particularly on DPUs \cite{AWSNitro2024,BlueFieldDPU_Blog,RN26}. As shown in Figure~\ref{fig:dpu}, the frontend \texttt{vring} on the compute node and the backend \texttt{shadow vring} on the DPU exchange data via DMA. The DPU’s \texttt{vring} connects to corresponding polling DPs through DMA, with DPDK or SPDK interacting with the hardware NIC via DMA. Numerous DMA operations occur along the I/O path, and DPDK/SPDK continuously poll DPs for fast data processing, which is critical for high DPU performance. To support this, \textbf{DPU CPUs are primarily used for polling data-plane requests, with only a few reserved for control-plane processes,} and \textbf{the high-speed I/O path’s multiple data transfers require extensive DMA operations and large amounts of memory involved in DMA.}

Moreover, the data plane of DPDK and SPDK extensively uses \texttt{hugetlb} and requires it to be as contiguous as possible to improve address translation efficiency. For network and storage services, substantial memory is also allocated to control-plane data, including network functions such as \texttt{vports} and \texttt{sessions} for access control, rate limiting, traffic statistics, and fast forwarding, and storage functions such as \texttt{queues} for device and queue management, I/O error recovery, second-level monitoring, and traffic statistics. For performance reasons, memory used by the data plane and control plane relies heavily on \texttt{hugetlb}. The amount of such memory depends heavily on the DPU’s supported specifications—the higher the service specification, the greater the memory consumption. However, overall DPU resources are limited; for example, early hardware had only 16\,GB of memory, and mainstream versions now have 32\,GB. \textbf{As product specifications grow, DPU resource usage remains tight even with hardware upgrades.}

\subsubsection{Static Memory Usage}

\begin{figure}[h!]
    \centering
    \includegraphics[width=0.9\linewidth]{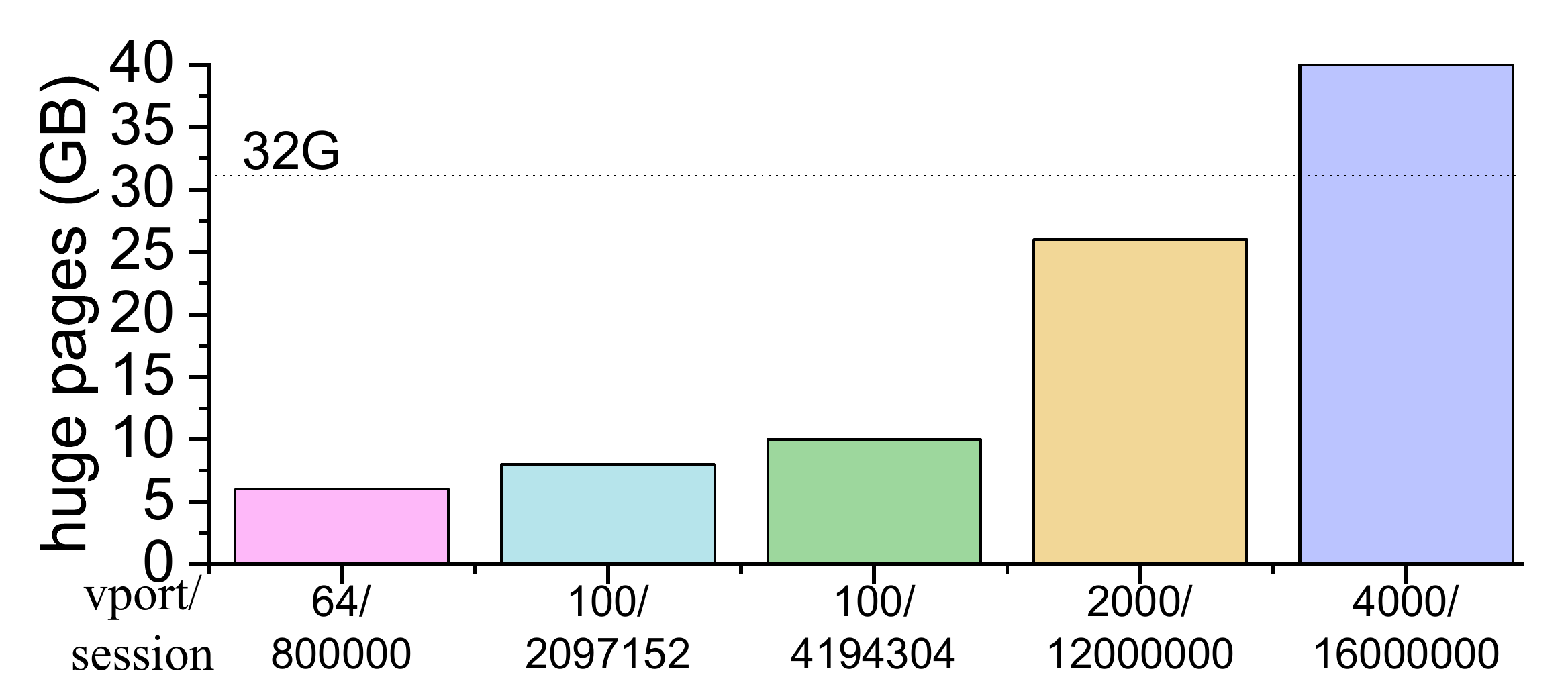}
    \caption{The DPDK Huge Page Usage}
    \label{fig:dpdk_hpage}
\end{figure}

While total memory consumption is complex and involves various data structures, memory usage per control data type in the same version can be approximated as \texttt{f(n)} = \texttt{base} + \texttt{each} × \texttt{n}, where \texttt{n} is the number of instances (e.g., \texttt{vport}, \texttt{session}, \texttt{queue}), \texttt{base} is the base memory to enable the function, and \texttt{each} is the per-instance overhead—e.g., each storage \texttt{queue} uses $\sim$2.7\,MB, and each network requires $\sim$1\,GB per 2 million sessions. This overhead may vary across versions due to added features or optimizations. As shown in Figure~\ref{fig:dpdk_hpage}, DPDK reserves memory for service metrics, with consumption increasing as specifications grow. Resources are often preallocated to peak levels to handle workload spikes and avoid allocation failures from fragmentation, potentially causing waste. Moreover, during hot upgrades, \texttt{dpdk} and \texttt{spdk} require data migration between old and new processes, necessitating an additional equal amount of memory reserved in advance (currently 8\,GB on a 32\,GB DPU). Thus, \textbf{to support maximum product specifications, storage and network services preallocate large amounts of statically allocated memory, with an additional equal amount reserved for hot upgrades.}

\subsubsection{Dynamic Memory Usage}
\begin{figure}[h!]
    \centering
    \begin{subfigure}[b]{0.48\linewidth}
        \centering
        \includegraphics[width=\linewidth,height=3cm]{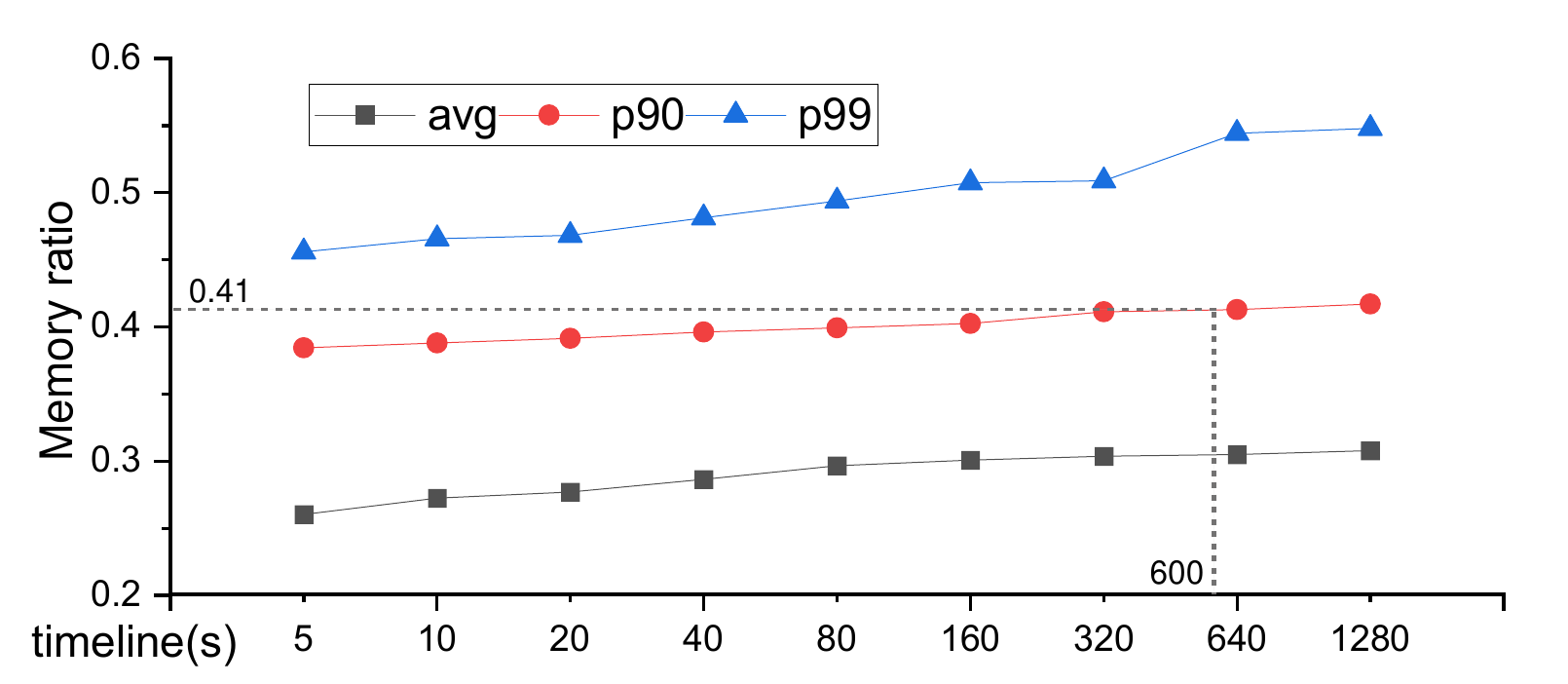}
        \caption{The DPU Hot Data Ratio}
        \label{fig:hot_ratio}
    \end{subfigure}
    \hfill
    \begin{subfigure}[b]{0.48\linewidth}
        \centering
        \includegraphics[width=\linewidth,height=3cm]{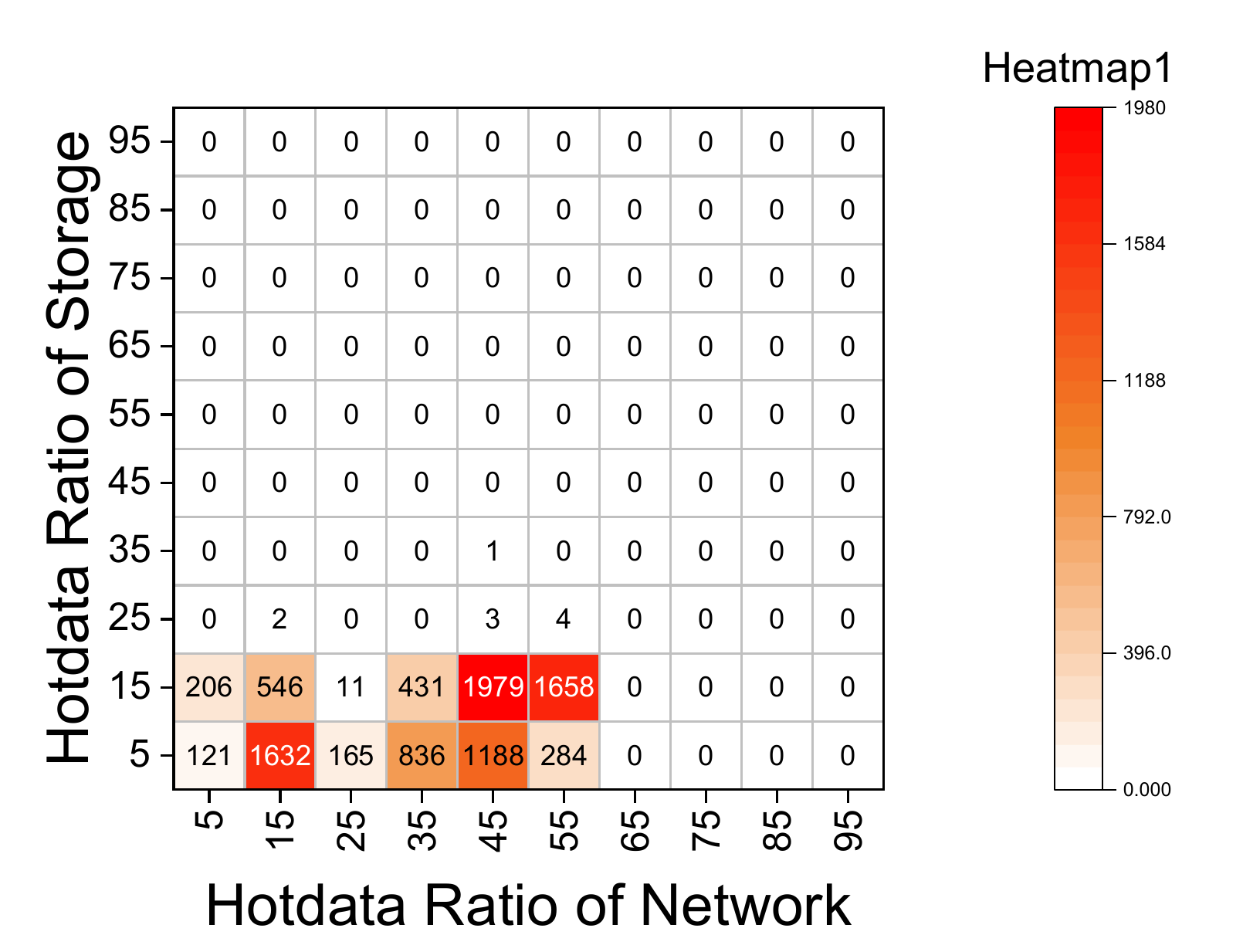}
        \caption{The DPU Hot Data Hotmap}
        \label{fig:hot_both}
    \end{subfigure}
    \caption{Comparison of DPU hot data metrics.}
    \label{fig:hot_comparison}
\end{figure}

As shown in Figure~\ref{fig:hot_ratio}, although much memory is statically allocated in huge pages to DPDK and SPDK for peak performance, most remains idle: the 10-minute average cold page ratio exceeds 70\%, and P90 exceeds 50\%. As shown in Figure~\ref{fig:hot_both}, DPDK and SPDK rarely peak simultaneously. Based on one-hour data from over 7,000 servers, both are rarely busy—network activity is mostly below 60\%, storage below 20\%—with no observed simultaneous peaks. \textbf{Under real workload pressure, the utilization of statically allocated memory is low, with many cold pages, and storage and network services rarely reach peak performance simultaneously, with only occasional single-side high load.}

\captionsetup[figure]{skip=3pt}

\noindent\framebox[\linewidth]{\parbox{0.97\linewidth}{\textit{\textbf{Finding~1}: A large amount of huge page on the DPU, reserved for peak performance, remains unused in practice.}}}

\subsection{Challenges in Existing Solutions}
\label{Existing Solutions}
\subsubsection{Functionality and Efficiency Challenges with Memory Swapping}
\label{swap_solution}
Memory usage patterns in DPUs are closely tied to running services, making memory utilization changes difficult. For example, storage and network services often use huge pages like HugeTLB for high performance, with our online data frequently exceeding 75\%. Traditional memory swapping is rarely implemented at scale and mainly targets small pages \cite{zhang2017memflex}, making huge page swapping impractical. While swapping techniques have evolved to support THP \cite{fichte2020towards}, the predominant use of HugeTLB on DPUs—which remains unswappable—results in over 75\% of memory being non-swappable. Using traditional swapping methods limits expandable memory resources, failing to meet the expansion needs of business software on the DPU. Even if the OS is modified to support HugeTLB swapping, many kernel memory areas, such as slab and crashkernel, remain unswappable, preventing full use of cold memory.

Significant latency affects current memory swapping techniques. Even with memory backend, zswap \cite{weiner2022tmo,lagar2019software} incurs 40\textmu s average latency for 4KB pages. Swapping 2MB huge pages typically takes milliseconds \cite{wang2023efficient,guo2024vpri}. Failed huge page allocation triggers prolonged reclamation, further increasing latency. Fallback to small pages during swap severely degrades performance. Thus, existing solutions fail to meet DPU requirements \cite{mcclure2022efficient}, especially for storage and network services requiring 90\% swap-in latency under 10\textmu s. \textbf{A high-performance memory swapping solution is essential to preserve huge page benefits while meeting strict latency targets.}

\begin{table}[htbp]
\centering
\caption{Comparison of Memory Elasticity Solutions}
\renewcommand{\arraystretch}{1.7}

\tiny

\begin{tabular}{|p{0.75cm}|p{0.75cm}|p{1cm}|p{1.05cm}|p{0.95cm}|p{0.5cm}|p{1.15cm}|}
\hline
\multicolumn{2}{|p{1.5cm}|}{\textbf{Memory Elasticity Solution}} &
\textbf{Swappable Memory} & \textbf{Unswappable Memory} &
\textbf{Resource Overhead} & \textbf{Open Source} & \textbf{Swapping Mechanism} \\
\hline
\multirow{2}{1.1cm}{\textbf{Memory Swapping}} &
Traditional swap & User-space small pages &
Various huge pages, kernel pages & Low & Yes & zram, ksm, swap \\ \cline{2-7}
& THP swap & User-space small pages, THP &
HugeTLB, kernel pages & Low & Yes & THP huge pages \\ \hline

\multirow{4}{2cm}{\textbf{VM-Based \\ Solutions}} &
KVM & Full & None & High & Yes & \multirow{4}{1.15cm}{\raggedright
    \textbf{Host-level}: Rely on host \\ OS or hypervisor.\\
    \textbf{Guest-level}: Rely on guest \\ OS.\\
    \textbf{PV}: Balloon.
} \\ \cline{2-6}
& Xen & Full & None & High & Yes & \\ \cline{2-6}
& Hyper-V & Full & None & High & No & \\ \cline{2-6}
& VMware ESXi & Full & None & Low & No & \\ \hline
\end{tabular}
\end{table}
\subsubsection{Lightweight and Adaptability Challenges with Virtualization}
Virtualization enables resource elasticity \cite{pearce2013virtualization} via full VM memory swapping. However, mainstream Type 2 KVM \cite{zhang2020giantvm} requires a Host OS, incurring fixed overhead on a 32GB DPU—at least 524MB for struct page, 256MB for crash kernel, and over 600MB for the OS—about 5\% of memory, excluding virtualization overhead. Running two OSes doubles stability and security risks \cite{zhang2011cloudvisor}. Type 1 systems like Xen \cite{barham2003xen} and Hyper-V \cite{velte2009microsoft} require a privileged VM, incurring similar costs. VMware ESXi \cite{haletky2011vmware} avoids this but is closed-source and not directly usable. These approaches rely on Host OS or hypervisor swapping for full Guest OS memory swap, facing the same performance issues.

Moreover, DPU services rely on stable DMA operations and direct hardware interaction, making seamless migration to VMs difficult. Adapting them for VM execution requires significant modifications, leading to high adaptation costs and potential stability issues. Running these services in VMs may lengthen I/O paths, degrade performance, and require extensive optimization. Operating two OSes also increases maintenance overhead. Thus, maintaining dual OS instances raises operational complexity, stability risks, and security vulnerabilities. \textbf{Deploying traditional virtualization on resource-constrained DPUs incurs prohibitive costs, necessitating a lightweight virtualization approach tailored for DPU resource elasticity.}

\noindent\framebox[\linewidth]{\parbox{0.97\linewidth}{\textit{\textbf{Finding~2}: Current memory elasticity solutions cannot be directly applied to DPU environments due to limitations in functionality, performance, or resource overhead.}}}

\subsection{Challenges in Production Environments}
\subsubsection{Issues with Existing Online Deployments}
Our public cloud has deployed hundreds of thousands of DPUs, supporting many businesses that need resource elasticity to take advantage of the latest software updates and higher product specifications. Traditional memory swapping and virtualization solutions \cite{kivity2007kvm} cannot ensure transparency for online businesses and often require a DPU reboot to enable resource elasticity, disrupting user service continuity. While live migration can reduce impacts on businesses, the cost and duration of large-scale rolling upgrades are high, and it still does not offer a seamless experience for users \cite{clark2005live}, especially those with high-load and sensitive requirements, leading to potential service jitter.

Another key scenario for DPUs is providing secondary virtualization for bare metal servers \cite{BM07}. Currently, there are no effective solutions for live migration of bare metal instances, which represent over 30\% of online deployments and often demand higher performance. Additionally, public clouds have strict SLO guarantees for stable and reliable services, making large-scale reboots impractical for implementing resource elasticity. Thus, solutions requiring DPU reboots cannot be deployed on existing servers. \textbf{A seamless hot-deployment resource elasticity solution is needed to fully leverage resource elasticity benefits.}

\subsubsection{Maintainability and Extensibility of Elasticity Solutions}
In cloud computing, every component requires upgrades and maintenance, making hot upgrade capability essential \cite{zhang2019fast}. Without it, even a resource elasticity component that supports seamless deployment for existing and incremental DPUs will face challenges during future upgrades that affect users. Hot upgrades help resolve faults exposed by large-scale operations and allow for easy extension of new features (like CPU elasticity), enabling rapid release and iteration of the resource elasticity component in small, quick steps. Thus, \textbf{our resource elasticity architecture must include hot upgrade capabilities to quickly realize and expand the benefits of resource elasticity.}

Moreover, using separate solutions for incremental and existing DPU deployments increases maintenance costs and risks to online stability. Meanwhile, to ensure high performance, DPUs allocate most CPU resources to data-plane tasks, leaving limited capacity for control-plane tasks, which may become constrained as service scales grow. Traditional CPU elasticity methods often require OS scheduler modifications \cite{humphries2021ghost,thiyyakat2020improving,miller2024enoki}, leading to maintenance challenges similar to those in memory elasticity. To support future upgrades and maintenance, \textbf{we seek a unified virtualization-based resource elasticity solution that minimizes components and operational complexity}, serving both new and existing DPUs while addressing immediate memory elasticity and enabling future CPU elasticity.

\noindent\framebox[\linewidth]{\parbox{0.97\linewidth}{\textit{\textbf{Finding~3}: Fully realizing the benefits of resource elasticity requires hot-switch and hot-upgrade capabilities for large-scale deployment on both existing and new DPUs.}}}

\section{Overview}
\subsection{Objectives}
Based on the above analysis, DPUs statically reserve substantial resources for peak demands that often remain idle, presenting an opportunity for elasticity. Meanwhile, given limitations of existing solutions and production maintenance challenges, elasticity must meet the following requirements:

\textbf{O1 - Lightweight}: DPU resources are extremely constrained, so the elasticity solution’s overhead must be minimal—around 1\% of DPU resources—and must not degrade the DPU’s native performance, especially ensuring less than 5\% peak performance loss for storage and network services.

\textbf{O2 - High Efficiency}: The solution must be highly efficient with minimal impact on existing services; in particular, to avoid degrading high-performance I/O, the 90th percentile page fault latency must be under 10\textmu s; additionally, due to extensive DMA usage, memory elasticity must not introduce data correctness risks.

\textbf{O3 - High Benefit}: Given the DPU’s limited resources, the elasticity solution must deliver substantial resource gains—we target over 50\% memory elasticity—while supporting both routine operations and sudden high-load workloads.

\textbf{O4 - Transparency}: To accommodate large-scale online DPUs, the solution must enable transparent deployment without disrupting existing services, and support transparent upgrades for seamless large-scale rollout.

\subsection{Architecture}

\begin{figure}[h!]
    \centering
    \includegraphics[width=0.9\linewidth]{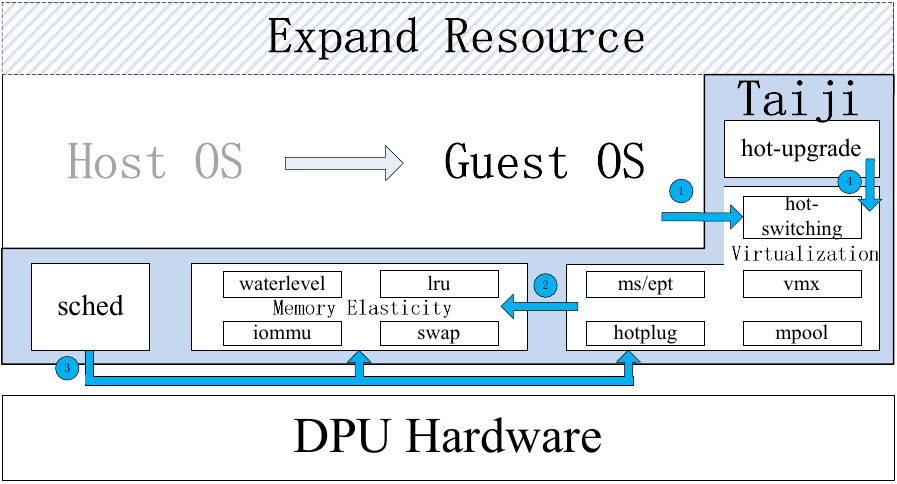}
    \caption{The Architecture of Taiji.}
    \label{fig:arch}
\end{figure}

As shown in Figure \ref{fig:arch}, Taiji consists of four main modules: the Virtualization Layer enables basic virtualization functions, Memory Elasticity provides resource expansion, the Scheduler coordinates Guest OS and elastic task execution, and Hot Upgrade supports online upgrades. The main functionality of each module are as follows:

\subsubsection{Virtualization Layer}
Any resource elasticity solution must not worsen constraints, especially on memory-limited DPUs. DPU virtualization here does not manage multiple VMs—it only virtualizes the existing OS to overcome physical limits. To avoid OS overhead, Taiji hot-switches the DPU’s OS online (\textbf{O4}) to insert a minimal virtualization layer, achieving lightweight virtualization (\textbf{O1}) and enabling resource elasticity (\bluecircled{1}). After switching, the Host OS becomes a Guest OS, and Taiji controls the hardware. From the virtualization layer’s view, all Guest OS memory becomes swappable, enabling memory elasticity. Virtual memory is then hot-plugged into the Guest OS, allowing time-division multiplexing of physical memory (\textbf{O3}).

\subsubsection{Memory Elasticity}
Due to high-performance DPU service requirements, memory elasticity must be efficient (\textbf{O2}) without performance loss. Taiji tracks physical memory hot/cold status using a multi-level LRU and multi-task concurrency to speed data collection. It proactively swaps cold pages to backend storage via parallel SWAP to enable Guest OS memory elasticity (\bluecircled{2}). A concurrent swap-in design keeps P90 swap-in latency under 10\textmu s. As much DPU memory may be involved in DMA, the swapping process avoids DMA memory. A watermark-based policy controls swapping to prevent thrashing, reserve free memory for sudden spikes, and ensure stable operation (\textbf{O3}).

\subsubsection{Resource Scheduling}
With memory elasticity enabled, Taiji runs both switched front-end VCPU and background elasticity tasks. To prevent background tasks from affecting front-end services, especially on DPs, a scheduler coordinates execution (\bluecircled{3}). Based on DPU CPU usage, CPs let original services yield time slices to elasticity tasks. Multi-task memory elasticity also requires parallel scheduling to meet \textbf{O2}. The scheduler coordinates execution using static configuration and dynamic adjustment: assigning priorities, allocating time slices, and adjusting them based on runtime feedback to ensure fair scheduling across priorities.

\subsubsection{Hot Upgrade}
To maximize DPU resource elasticity, Taiji is designed to support large-scale existing online DPUs from the outset. Its virtualization layer enables hot-switch for seamless online transition to the elastic architecture. As a core DPU component, Taiji also supports hot-upgrade (\bluecircled{4}) through a modular, compatibility-driven design for future updates and bug fixes. With hot-switch and hot-upgrade (\textbf{O4}), Taiji can run on all existing and new DPUs as a unified resource management platform without disrupting running services, enabling large-scale production rollout.

\section{Design and Implement}

\subsection{Virtualization Layer}

\subsubsection{Lightweight Hybrid Virtualizationl}
To achieve \textbf{O1}—a low-overhead lightweight virtualization solution—Taiji designs the virtualization layer to reduce both resource and performance overhead. \textbf{Performance overhead} is minimized using register passthrough and huge page support. Taiji’s Virtual Machine Control Structure (VMCS) design applies register passthrough whenever possible to avoid costly VM exits, removing most virtualization overhead \cite{virtperf, li2021accelerator}. As huge pages are common in DPUs, the virtualization layer’s memory management retains this feature to avoid address translation overhead, with the underlying extended page table (EPT) using huge pages to reduce TLB miss impact \cite{agarwal2017thermostat, jia2023making, bergman2023translation}. \textbf{Resource overhead} reduction requires a new, more complex hybrid virtualization architecture. As shown in Figure~\ref{fig:virt}, Taiji switches physical CPUs (PCPUs) to virtual CPUs (VCPUs), hot-switch the original Host OS into a Guest OS and inserting a thin virtualization layer. The switched Host OS runs in non-root mode as the Guest OS, while the Taiji module resides in the Guest OS but operates mainly in root mode to manage hardware (Host layer), with a small interface layer in non-root mode (Guest layer). This design takes over hardware resource management while reusing the Guest OS user space, avoiding the need for an additional user layer, and forms a hybrid virtualization architecture distinct from traditional Type 1 and Type 2 designs. Specifically, it has the following particularities:

\begin{figure}[h!]
    \centering
    \includegraphics[width=\linewidth]{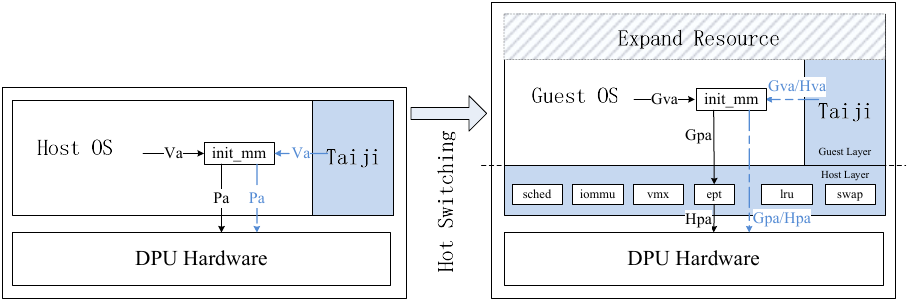}
    \caption{The Hybrid Virtualization Architecture}
    \label{fig:virt}
\end{figure}

\textbf{First}, Taiji is implemented as a module inserted into the OS to run both before and after the virtualization switch. Before the switch, Taiji functions as a regular kernel module. After the switch, the Host OS becomes the Guest OS, but Taiji remains a module in the Guest OS while taking over the underlying hardware. Most of Taiji runs in root mode, with a small interface part in non-root mode, and both share the Guest OS’s user space.

\textbf{Second}, since the Taiji module runs inside the OS kernel, they share the same kernel address space and thus the same virtual addresses. Before the switch, both use the OS virtual address (VA). After the switch, the OS uses the Guest Virtual Address (GVA) in non-root mode, and Taiji uses the Host Virtual Address (HVA) in root mode. Sharing the same kernel address space results in GVA = HVA.

\textbf{Third}, because Taiji runs inside the OS, its address translation also uses the kernel page table (\texttt{init\_mm}) to convert VA to PA. Before the switch, this is identical to the Host OS. After the switch, memory access paths diverge: when the Guest OS accesses memory, its GVA is first translated via \texttt{init\_mm} into a Guest Physical Address(GPA), which is then converted by EPT into an Host Virtual Address (HPA) to access physical memory.

\textbf{Fourth}, after the switch, when the Taiji module accesses physical memory, it still uses \texttt{init\_mm} to translate its VA into a GPA. The interface part running in the Guest passes through EPT to obtain the HPA. However, since most of the Taiji module runs in root mode, the main virtualization layer bypasses EPT and directly accesses physical memory via GPA. This single-layer page table translation improves memory access performance \cite{zhang2024direct} but requires GPA = HPA to ensure correctness.

\textbf{The most distinctive feature of this virtualization is its address space management.} Running inside the Guest OS while acting as a hypervisor, the Taiji module and Guest OS have \textbf{GVA = HVA}, fundamentally different from traditional architectures. Since Taiji accesses physical memory through a single-layer page table, correctness requires all \textbf{Taiji} accessed memory be \textbf{GPA = HPA}. Based on this, we design a metadata pool (\texttt{mpool}) that allocates full pages and slab memory at various granularities. All Taiji metadata is allocated from this pool, whose memory is pinned and excluded from swapping, ensuring GPA = HPA and forming the basis of the hybrid virtualization architecture. Centralized metadata management also prevents fragmentation, avoiding negative effects on future memory elasticity. Other Taiji functions are built on this pool.

\subsubsection{Virtualization Switching}
To achieve \textbf{O4}, i.e., deploying Taiji on running DPUs, the Host OS is hot-switched into a Guest OS by converting each PCPU to a VCPU. For every running PCPU, an EPT table, a VMCS, and other virtualization structures are prepared to minimize overhead. The switch is triggered by an \texttt{SMP} call to enter virtualization mode. Upon receiving the call, the target CPU executes \texttt{switch\_vcpu} in two stages. First, the PCPU saves registers, switches to the VCPU stack, transfers the saved state into the VMCS (\bluecircled{1}), and calls \texttt{hv\_sched}, which invokes \texttt{vcpu\_run}. This loads the VMCS and registers, then executes \texttt{VMLAUNCH} to enter non-root mode. The VCPU’s first instruction re-invokes \texttt{switch\_vcpu}, which runs the second stage, restores the stack, and completes the switch (\bluecircled{2}). The VCPU then resumes from the PCPU’s execution flow, returning from the \texttt{SMP} call. The Guest OS CPU continues as a VCPU until a privileged instruction or exit event triggers a VM exit (\bluecircled{3}), while the PCPU in root mode runs \texttt{hv\_sched} to schedule VCPUs and elastic memory tasks (Section~\ref{sched}, \bluecircled{4}). This process repeats for all PCPUs until the switch is complete, enabling a hot-switch to the virtualization architecture.

\begin{figure}[h!]
    \centering
    \includegraphics[width=\linewidth]{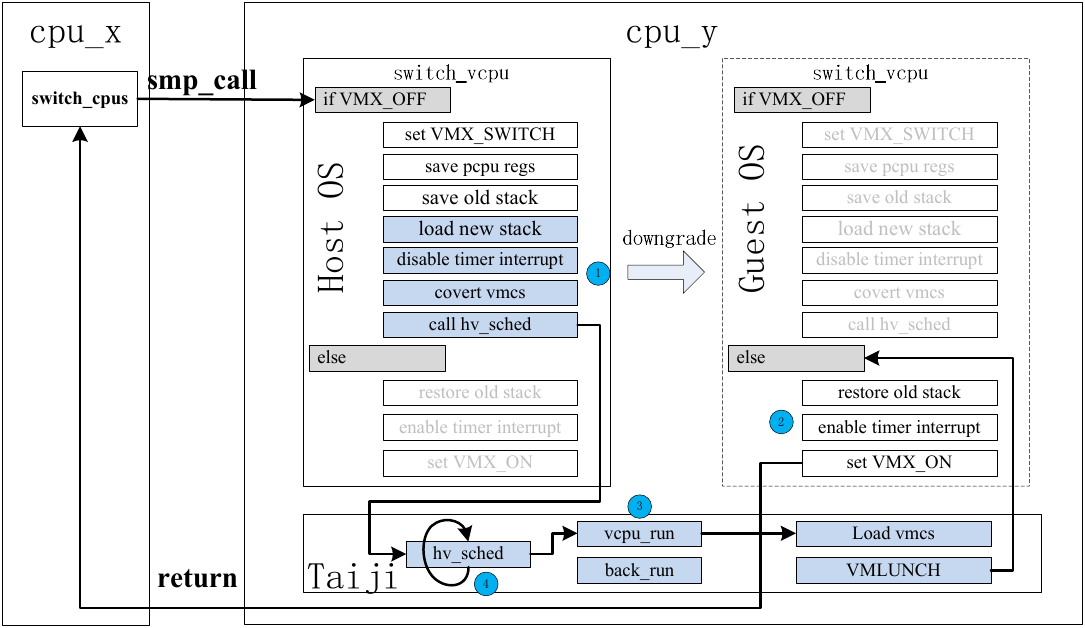}
    \caption{The VMX Switching Flow}
    \label{fig:vmx}
\end{figure}

\subsection{Memory Elasticity}

\subsubsection{Parallel Multi-Level LRU}
\begin{figure}[h!]
    \centering
    \includegraphics[width=\linewidth, height=3.5cm]{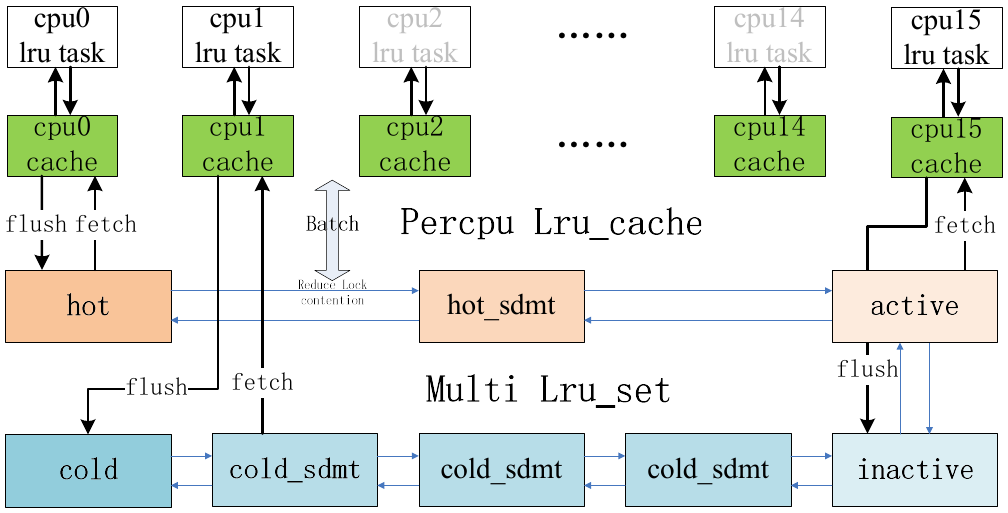}
    \caption{The Multi-level Hot/Cold Sets}
    \label{fig:lru}
\end{figure}

To achieve \textbf{O2} of elastic memory, accurate identification of hot and cold states of swapped memory is required. Taiji follows the huge page usage characteristics of the DPU and manages physical memory mainly at huge page granularity. The current kernel lacks an LRU mechanism for huge pages \cite{LinuxDoc2025MGLRU, Park2025DAMON, RN17}. Due to the coarse granularity, access to a single base page within a huge page may cause abrupt, transient changes in its hot/cold state that do not reflect actual thermal behavior. To address this, Taiji leverages temporal locality with time-based stabilization to smooth such fluctuations. It employs a multi-level hot/cold set structure with parallel multi-tasking for efficient, accurate cold page identification at huge page granularity. As shown in Figure \ref{fig:lru}, hot pages are kept in the \texttt{hot} set and cold pages in the \texttt{cold} set. Pages transitioning between hot and cold are placed in \texttt{inactive} and \texttt{active} sets. Intermediate sets between \texttt{hot} and \texttt{active}, and between \texttt{inactive} and \texttt{cold}, smooth stabilization during periodic scans; if a page’s state remains unchanged after a scan, it gradually shifts toward the \texttt{hot} or \texttt{cold} end. Within each set, elements are ordered by arrival time to indicate relative temperature, e.g., in the \texttt{cold} set, the head is colder than the tail. An LRU background task runs per PCPU to periodically scan and manage these sets, accelerating state updates. Each PCPU also maintains a scan cache to buffer scan tasks and results, reducing lock contention. The scheduler ensures LRU tasks execute timely to capture hot/cold state changes.

\subsubsection{Parallel Low-Latency SWAP}
The prerequisite for DPU memory elasticity is preserving native performance, requiring 90\% of page fault latencies $<$10\,$\mu$s (\textbf{O2}). Swap-path latency has two parts: 1) The \textbf{backend} choice critically affects latency: disk or file-based backends cannot meet requirements \cite{jo2023toward}, while keeping swapped data in memory greatly reduces it. Given the prevalence of cold pages on DPUs and the maturity of backend techniques, Taiji uses in-memory zero pages \cite{song2016efficient} and compression \cite{weiner2022tmo}, prioritizing zero pages to minimize backend latency. 2) For \textbf{frontend} granularity, since the smallest fault unit is a small page, swapping in at huge page granularity causes high latency \cite{wang2023efficient,guo2024vpri}. However, huge pages reduce performance and metadata overhead from small-page management and avoid allocation failures from fragmentation \cite{fichte2020towards}. To balance this, Taiji manages swapping at memory section (MS, huge page) granularity but operates at memory page (MP, small page) granularity. A huge page is fully swapped only when all its small pages are swapped in or out. Leveraging swap-in/swap-out differences, Taiji parallelizes MP-level swap-ins to meet \textbf{O2} efficiency requirements, while swap-outs are sequential to simplify processing.

\begin{figure}[h!]
    \centering
    \includegraphics[width=\linewidth]{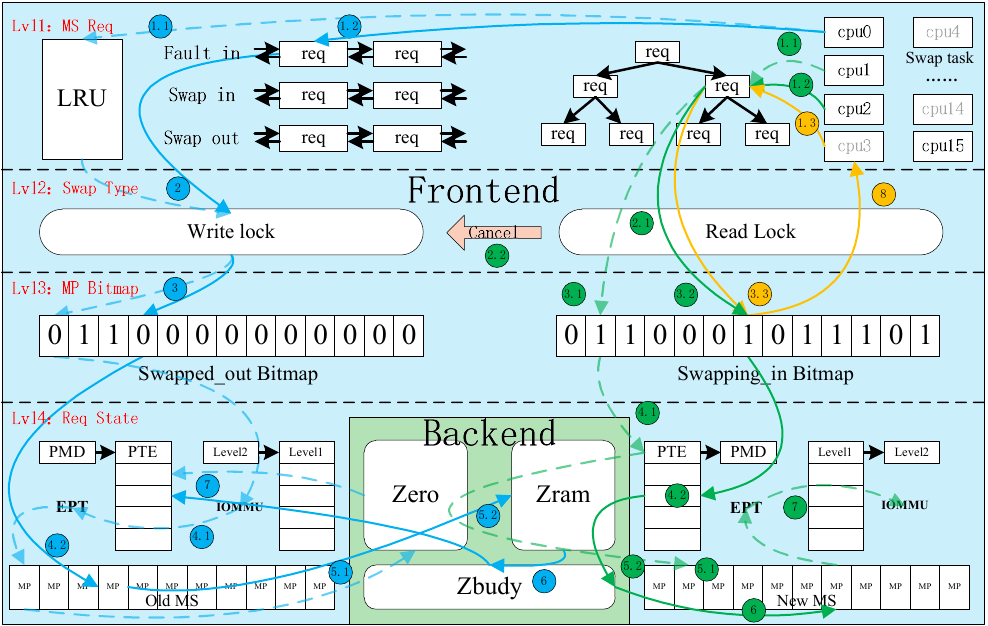}
    \caption{The Swap Parallel}
    \label{fig:swap}
\end{figure}

Thus, \textbf{controlling parallel swap-in of MPs within an MS is key to low-latency memory swapping}. Existing techniques lack parallel swap-in support for huge pages, requiring a design to ensure concurrency and atomicity of small-page swaps within them. Taiji defines an MS-level request entity \texttt{req} (containing GFN, PFN, locks, bitmaps, MS/MP state) for concurrency control, with three task types: \texttt{Fault\_in} (passive page-fault–triggered), \texttt{Swap\_out} (proactive reclamation), and \texttt{Swap\_in} (prefetch or compaction, omitted here). Concurrency control mainly uses the first two. As shown in Figure~\ref{fig:swap}, four atomicity layers—\textbf{req abstraction, read–write locks, execution bitmap, and MS/MP state}—coordinate to ensure efficient low-latency swap-in.

\textbf{First}, \texttt{req} operates at MS granularity, each with an independent MS-level lock to prevent contention across MSs and allow parallel swaps for different MSs. All \texttt{req}s are unique and stored in a red-black tree for efficient page-fault lookup. On initial swap-out, a new \texttt{req} is created from the LRU (\bluecircled{1.1}); for later active tasks, an existing \texttt{req} is fetched from the queue (\bluecircled{1.2}). Passive swap-ins find the corresponding \texttt{req} in the tree by the faulting address (\greencircled{1.1}–\greencircled{1.3}).

\textbf{Second}, the \texttt{req}’s read-write lock enforces mutual exclusion for active swap tasks while allowing passive swaps in parallel. Active tasks, without strict latency needs, are serialized via a write lock to simplify control flow (\bluecircled{2}). Passive swap-ins require low latency; since multiple MPs within an MS may fault concurrently (\greencircled{2.1}), read locks allow parallelism to reduce latency. On read-write lock conflicts, a cancel mechanism ensures the write-locked task exits promptly (\greencircled{2.2}).

\textbf{Third}, bitmaps ensure atomicity of swap operations on the same MP within an MS while enabling parallelism across MPs. Two bitmaps are maintained: an already-swapped-out bitmap and a currently-swapping-in bitmap. The first is set during swap-out (\bluecircled{3}) to ensure swap-in applies only to swapped-out MPs (\greencircled{3.1}–\greencircled{3.2}). The second guarantees atomicity for page fault–driven swap-ins: if multiple faults hit the same MP, only one proceeds with the swap-in (\orangecircled{3.3}).

\textbf{Finally}, state control prevents races during critical operations. For swap-out, EPT and IOMMU page tables are split at the first MP’s swap-out (\bluecircled{4.1}) and the MS is reclaimed after the last MP’s swap-out. For swap-in, a new MS is allocated at the first MP’s swap-in and EPT/IOMMU tables are merged after the last MP’s swap-in (\greencircled{7}). These operations must execute exactly once at defined state transitions, with parallel flows avoiding races that could cause data corruption.

After frontend concurrency and correctness control, swap-out and swap-in access backend storage for memory writes (\bluecircled{5}–\bluecircled{7}) or reads (\greencircled{4}–\greencircled{6}). These concurrency designs achieve high swap-in performance with low complexity. With coordinated frontend and backend designs, Taiji swaps in pages on EPT page faults within 10\textmu s at the 90th percentile, meeting \textbf{O2}. To meet \textbf{O3}, Taiji adopts a watermark-based swapping policy, allowing to reserve free memory for DPU memory bursts. Three watermarks are set: \texttt{high}, \texttt{low}, and \texttt{min}. Swapping starts when memory drops below \texttt{low} and stops when it rises above \texttt{high}. \texttt{min} marks critically low memory, triggering proactive swap-out during page faults to avoid prolonged low-memory states. Policies can be tuned, such as halting reclaim between \texttt{low} and \texttt{high} if no cold pages exist, or starting reclaim below \texttt{high} to prepare for sudden demand.

\subsection{Resource Scheduler}
\label{sched}
\begin{figure}[h!]
    \centering
    \includegraphics[width=\linewidth]{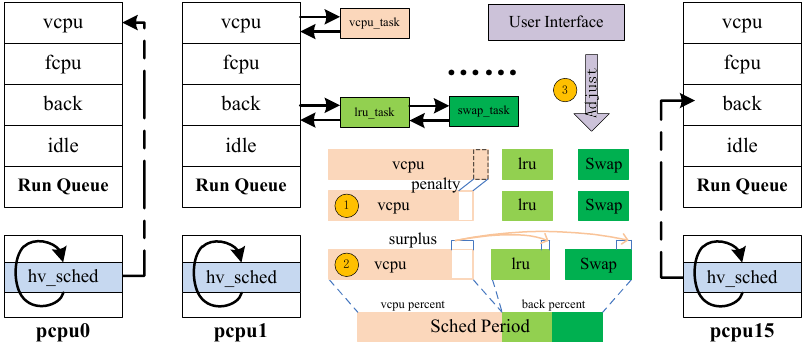}
    \caption{The Scheduler Architecture}
    \label{fig:sched}
\end{figure}

After resource elasticity, parallel elasticity tasks in Taiji consume CPU, but most DPU CPUs are dedicated to high-performance DPs, leaving only CPs for background tasks. CPs can yield resources to elasticity tasks but still require service guarantees. To avoid impacting native workloads, a scheduler coordinates VCPU tasks for Guest OS workloads and background memory-elasticity tasks. As shown in Figure~\ref{fig:sched}, once the virtualization layer activates and PCPUs are switched to VCPUs, each VCPU becomes a PCPU task managed by the \texttt{hv\_sched} scheduler in root mode. The scheduler runs in fixed cycles, adjusting background task time slices by CPU load and elasticity urgency, ensuring balanced execution through static configuration and dynamic adjustment.

First, the scheduler prioritizes VCPU execution, running background tasks only when resources permit. Since native DPU workloads must not be affected and prolonged VCPU starvation can hang the Guest OS, static configuration assigns priorities to task types. Each priority level receives a proportional share of execution time in each scheduling cycle based on task urgency. After a PCPU is virtualized, a run queue (\texttt{rq}) is created per PCPU with four priority queues: 1) VCPU, active VCPUs after switching; 2) FCPU, reserved for future new VCPU creation; 3) BACK, background tasks such as lru and swap; 4) IDLE, idle tasks. Higher-priority queues get larger execution shares, and tasks within each level share that level’s slice.  

Second, dynamically, the scheduler tracks each task’s start time and enforces a maximum duration. If a task exceeds its limit, a penalty reduces its time slice in later cycles (\orangecircled{1}). Within a cycle, unused time slices are reallocated to tasks of the same or lower priority (\orangecircled{2}). Users can adjust the set of CPs allowed for background tasks and their time slices via monitoring tools (\orangecircled{3}). At the end of each cycle or upon configuration updates, the scheduler recalculates allocations based on updated ratios and cycle period. Proportional allocation and fair adjustments ensure precision across priority levels, enabling efficient elasticity tasks without impacting native DPU workloads.

\subsection{Hot Upgrade}
To achieve \textbf{O4}, Taiji requires hot-upgrade capability in addition to hot-switch. This is challenging, as it must replace the running virtualization logic online. Once active, Taiji’s scheduler runs in a loop with prioritized tasks, each priority having its own operation set, and multiple external entry points bound to specific logic that may be invoked anytime. A hot upgrade must seamlessly redirect all old-module operations to the new module at the right moment without disrupting stability.

\begin{figure}[h!]
    \centering
    \includegraphics[width=0.9\linewidth,,height=3.5cm]{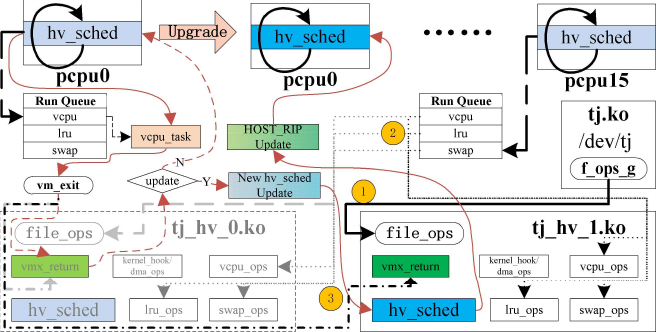}
    \caption{The Upgrade of Taiji.}
    \label{fig:upgrade}
\end{figure}

To this end, Taiji’s hot-upgrade solution is designed architecturally to allow updating by simply loading a new module. Based on this modular approach, Taiji is split into two independent parts: \texttt{tj.ko} (entry functionality) and \texttt{tj\_hv\_x.ko} (main functionality), as shown in Figure \ref{fig:upgrade}. The simple \texttt{tj.ko} logic does not require upgrading, while \texttt{tj\_hv\_x.ko}, containing complex logic, does. This allows the underlying complex functions to be upgraded transparently to user services relying on Taiji. The specific upgrade details are as follows:

\textbf{Data Plane Compatibility}: To simplify hot upgrades and avoid data conversion, Taiji’s metadata structures must remain compatible so that metadata managed by the old module can be directly inherited. Structure sizes must remain unchanged, with reserved fields for future use. Unless through complex conversion, the semantics and positions of existing fields must not change.

\textbf{Various Operation Entry Points}: The entry module (\texttt{tj.ko}) provides a unified file operation entry point. Through \texttt{\/dev\/tj}, each device operation is linked to a file structure whose \texttt{f\_ops} field points to the global entry \texttt{f\_ops\_g}, which redirects to actions in \texttt{tj\_hv\_x.ko}. For a hot upgrade, only this global entry needs updating, not every open file structure (\orangecircled{1}). Similarly, functional interfaces (e.g., per-priority scheduler operations, encapsulated \texttt{dma\_ops}, and externally registered functions) are updated to the new module’s addresses (\orangecircled{2}). All updates occur only after calls to the old module complete.

\textbf{VCPU Execution Transition}: After the VCPU switch, the PCPU cyclically executes the original module’s \texttt{hv\_sched}. During an upgrade, a VCPU may be running in non-root mode and unable to respond promptly to the upgrade request. On Exit, the VCPU returns to the old module’s exit location recorded in the \texttt{HOST\_RIP} field of the VMCS. To handle this, each PCPU maintains an update flag and the new module’s loop entry; if set, execution jumps to the new scheduler loop and updates \texttt{HOST\_RIP} to the new exit location, enabling a seamless handoff (\orangecircled{3}). When all VCPUs complete this switch, the Taiji hot upgrade is finished.

\section{Evaluation}
Taiji is implemented in the Linux~4.19 kernel, runs on Intel-based DPUs, and has been deployed at scale over 30{,}000 units. Performance results show no degradation of DPU performance, confirming virtualization efficiency (Section~\ref{perf}). Static and dynamic overhead evaluate its lightweight design for constrained DPUs (Section~\ref{light}). High-load hot-upgrade runtime data and online production statistics further demonstrate the efficiency and benefits of its elasticity (Section~\ref{elastic}).

\subsection{Performance}
\label{perf}
The performance test used two pairs of DPU cloud servers. The compute node had an AMD 2.7\,GHz 96-core CPU with 1\,TB memory. The DPU had an Intel(R) 2.00\,GHz 16-core CPU, with one group using a 32\,GB DPU plus 16\,GB virtual elastic memory, and the other a non-virtualized 64\,GB DPU.

\subsubsection{Benchmark}
\begin{figure}[t]
\begin{center}
\includegraphics[width=\linewidth,height=4cm]
{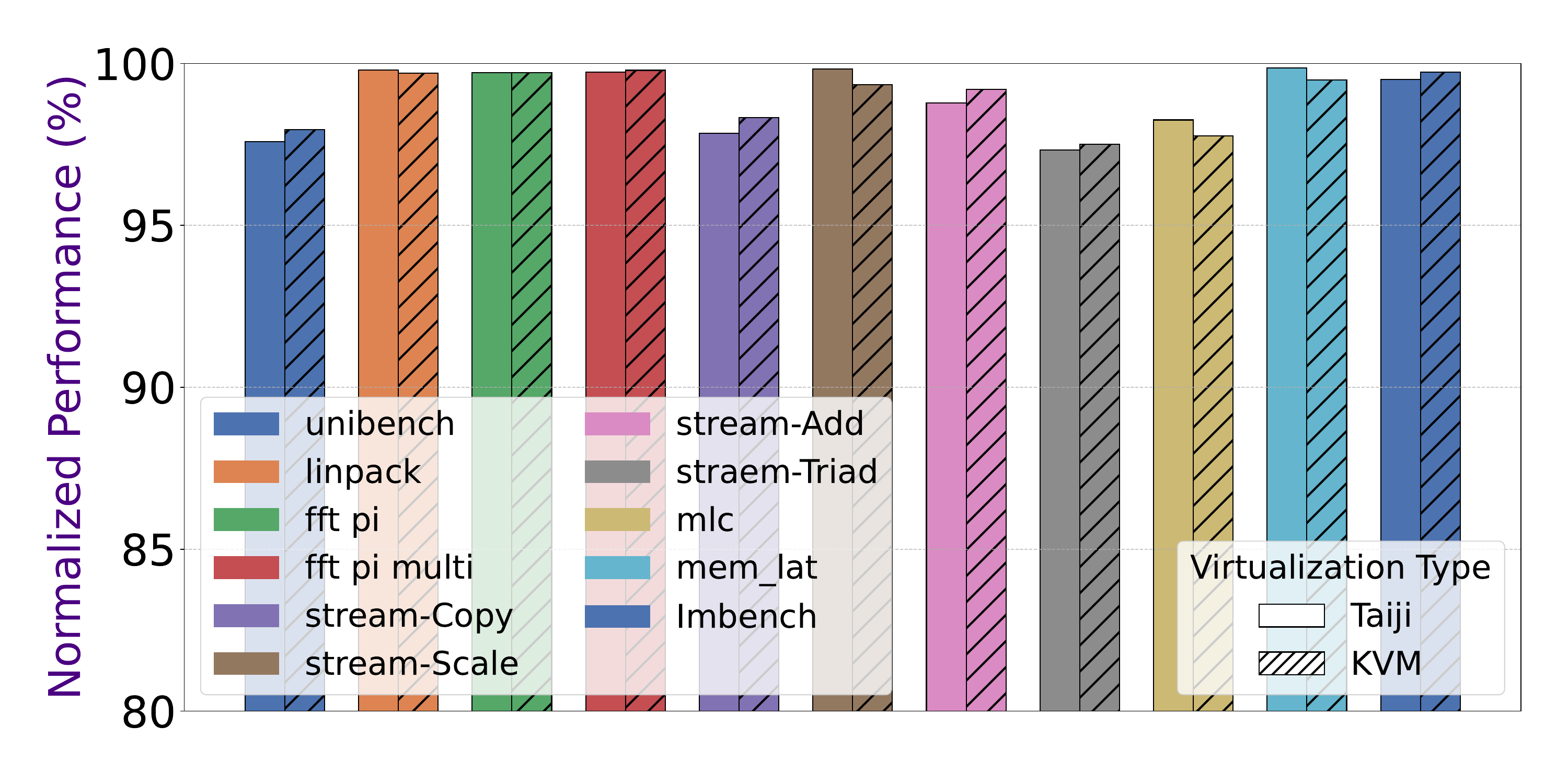}
\end{center}
\caption{The Normalized Performance of Benchmark.} 
\label{fig:benchmark}
\end{figure}

We employ a comprehensive set of benchmarks to evaluate Taiji’s virtualization overhead, covering CPU performance, memory bandwidth, latency, and cache management. To verify the virtualization efficiency of Taiji, we compared it with KVM. Figure~\ref{fig:benchmark} presents the results. Key observations: 1) Unibench: Taiji shows a 2.4\% drop, within normal virtualization overhead. 2) Linpack: Performance is nearly identical, overhead negligible. 3) FFTPI: Similar to Linpack, with only minor differences from fluctuation or experimental error. 4) MLC: 1.8\% difference, with stable results and minimal overhead. 5) Stream: maximum difference of 2.6\%, within acceptable limits. Overall, DPU CPU and memory performance under Taiji virtualization is comparable to native, with total overhead under 3\%. Compared to KVM virtualization, both incur similar performance loss. These results indicate minimal impact, meeting \textbf{O1}.

\subsubsection{Cloud Workload}
\begin{figure}[t]
\begin{center}
\includegraphics[width=\linewidth,height=4cm]{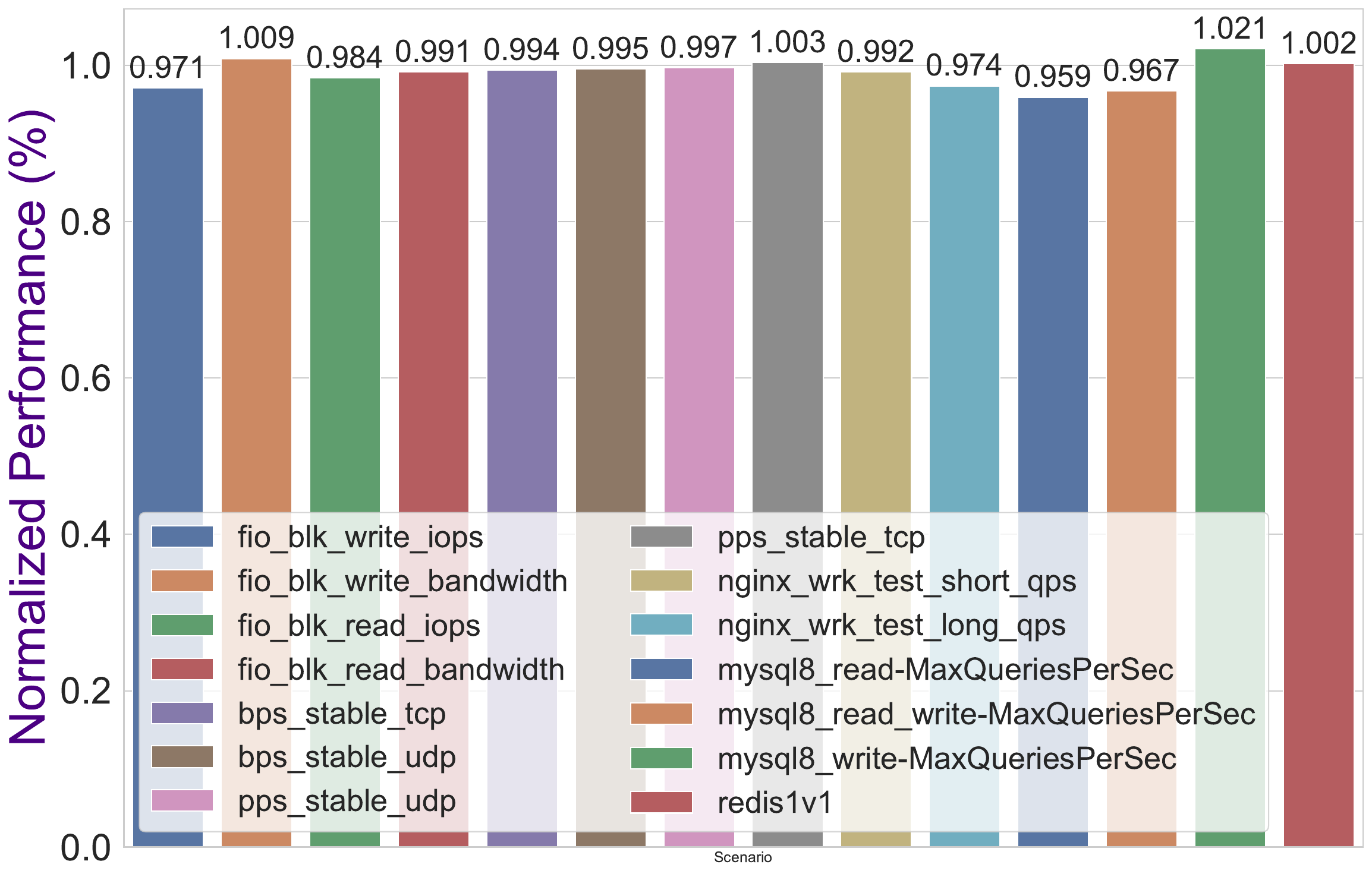}
\end{center}
\caption{The Normalized Performance of Cloud Workload.}
\label{fig:workload}
\end{figure}

Figure~\ref{fig:workload} evaluates Taiji’s impact on VM business performance using instance-selling workload tests. 1) Storage: In tests such as \texttt{fio\_blk}, Taiji achieves performance similar to native; IOPS drop by about 3\%, bandwidth is nearly identical, and overhead is within expectations. 2) Network: \texttt{bpps} and \texttt{pps} show minor fluctuations from random load variations, with performance ratio near 1.0. \texttt{nginx} drops slightly, within 3\%, indicating minimal loss. 3) MySQL and Redis: \texttt{mysql8\_read} throughput decreases slightly, within acceptable limits. Overall, compared to native, Taiji introduces no significant difference in core DPU storage and network performance, with memory elasticity overhead acceptable, demonstrating that its memory elasticity and task scheduling meet \textbf{O1}.

\subsection{Lightweightness}
\label{light}
\subsubsection{Code Size}

Table~\ref{tab:module_code_lines} shows the code size distribution of Taiji’s modules. All are lightweight; although core modules such as VMX, LRU, and Swap are slightly larger, they are far smaller than their Linux counterparts. VMX, the largest and most complex, has 9{,}557 LOC, compared to KVM’s 77{,}000 LOC. Memory management modules (MS: 3{,}273; LRU: 4{,}202; Swap: 4{,}101) are minimal relative to Linux’s 151{,}000. The scheduler (Sched) has 2{,}755 lines, versus Linux’s 42{,}000. Thus, unlike traditional virtualization that adds an entire OS, Taiji remains lightweight in code complexity and does not worsen DPU resource constraints.

\begin{table}[htbp]
\centering
\caption{Code Lines for Each Module in Taiji}
\footnotesize
\begin{tabular}{lccccccccc}
\toprule
\texttt{MOD} & \texttt{Mpool} & \texttt{MS} & \texttt{VMX} & \texttt{Attr} & \texttt{LRU} & \texttt{Sched} & \texttt{Swap} & \texttt{API} \\
\midrule
2567 & 2492 & 3273 & 9557 &3158 & 4202 & 2755 & 4101 & 3063 \\
\bottomrule
\end{tabular}
\label{tab:module_code_lines}
\end{table}

\subsubsection{Metadata}
\begin{figure}[htbp]
    \centering
    \begin{minipage}{0.23\textwidth}
        \centering
        \vspace{0.3cm} 
        \includegraphics[width=\linewidth,height=3cm]{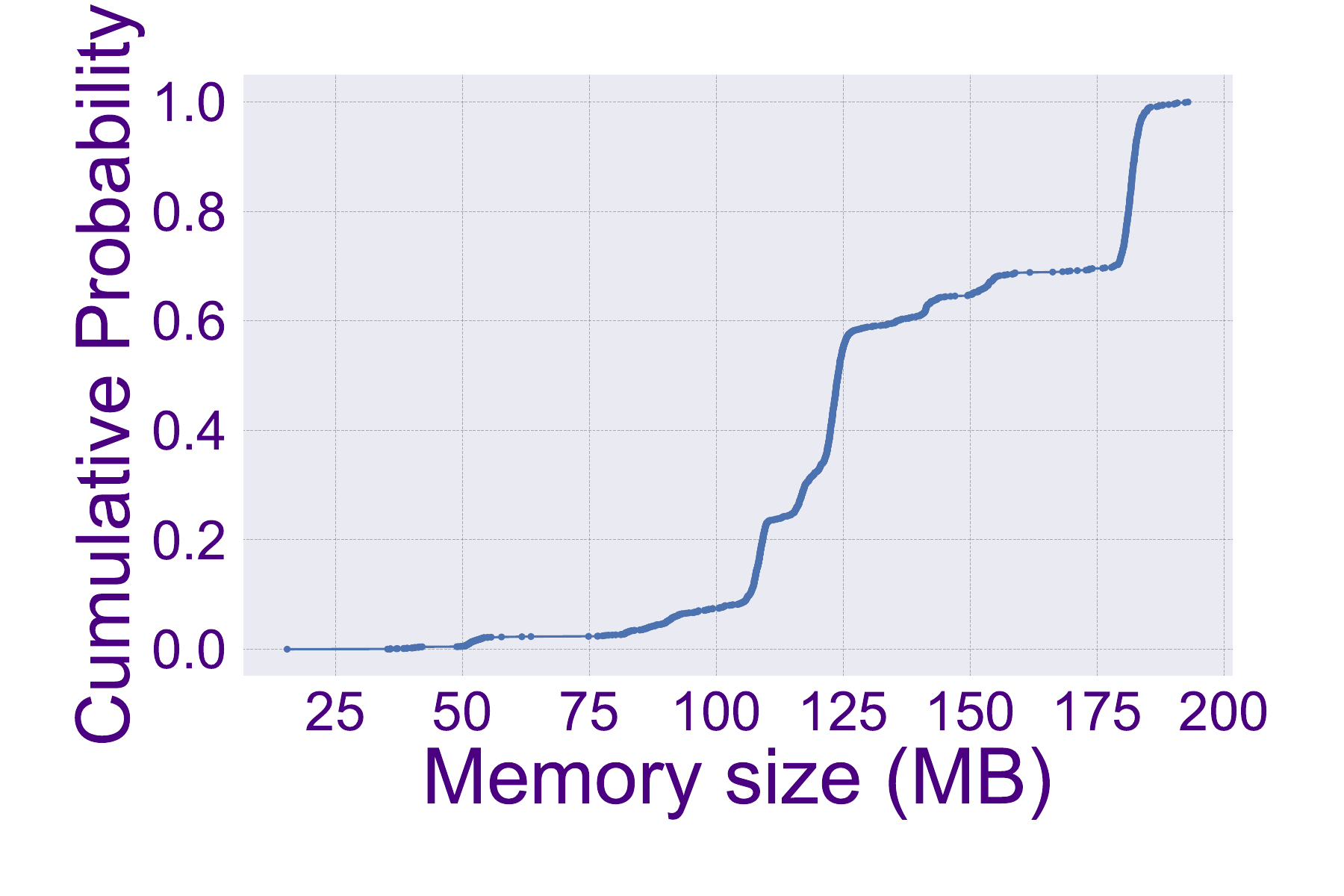}
        \subcaption{The Metadata Size.}
        \label{fig:meta}
    \end{minipage}\hfill
    \begin{minipage}{0.23\textwidth}
        \centering
        \includegraphics[width=\linewidth,height=3cm]{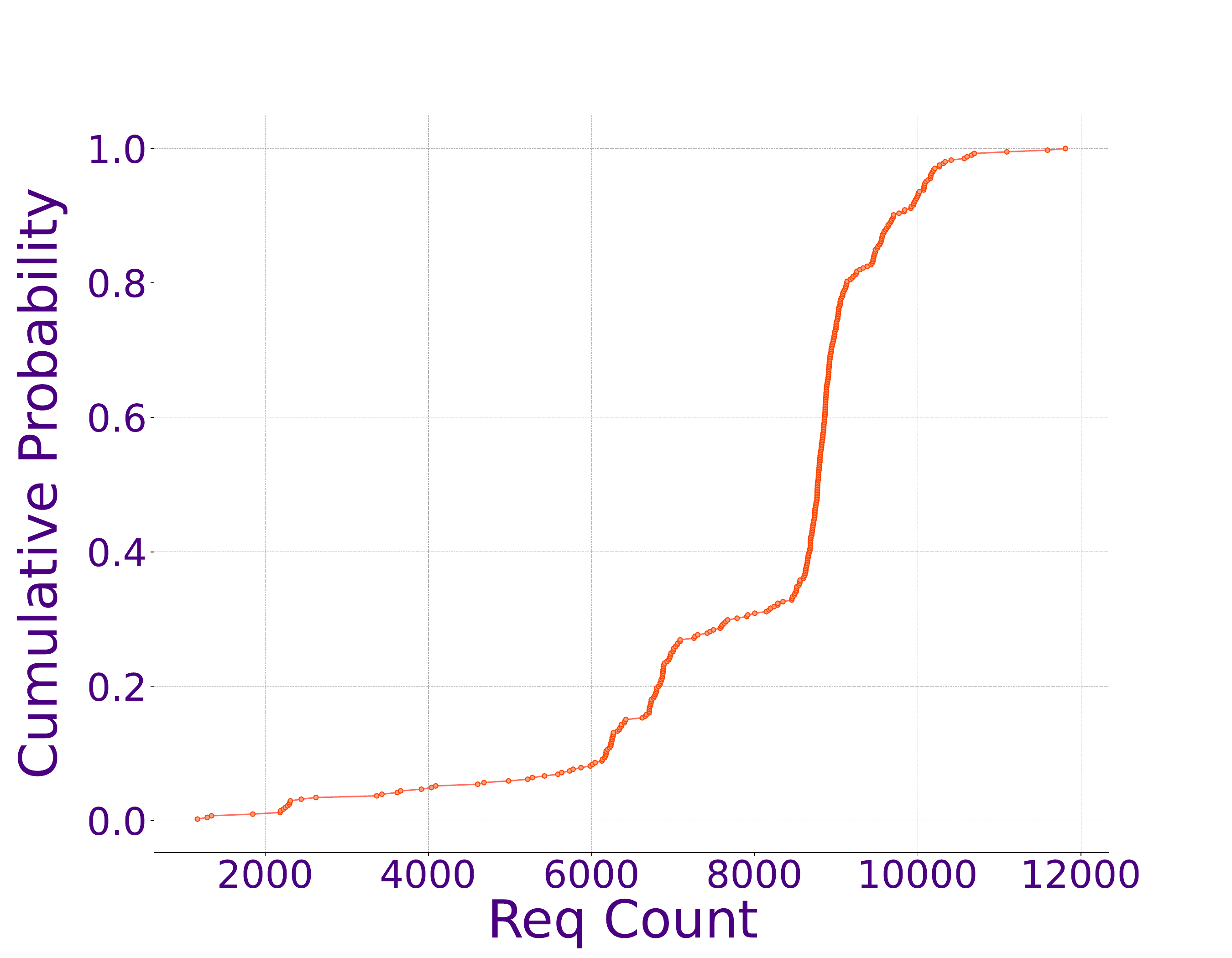}
        \\
        \subcaption{The Swap \textit{req}}
        \label{fig:req_cluster}
    \end{minipage}
    \caption{The Running Cost and Benefit of Taiji.}
    \label{fig:overall}
\end{figure}

Figure~\ref{fig:meta} shows the CMF of metadata utilization in Taiji’s \texttt{mpool}. Of the 400\,MB reserved, average usage is 127.33\,MB (46.69\%), peaking below 200\,MB; 68.53\% is for full pages (EPT and IOMMU page tables) and 31.47\% for slab pages (swap, LRU). By minimizing overhead, Taiji uses under one-third of reserved memory, enabling future expansion. Including all reserved metadata, total resource overhead is 1.2\% and actual metadata overhead 0.38\%, far below the $\sim$5\% of conventional virtualization, meeting \textbf{O1}.

\subsection{Memory Elasticity}
\label{elastic}

\subsubsection{High-Load Operation Analysis}
\begin{figure}[htbp]
    \centering
    \begin{minipage}{0.23\textwidth}
        \centering
        \includegraphics[width=\linewidth,height=3cm]{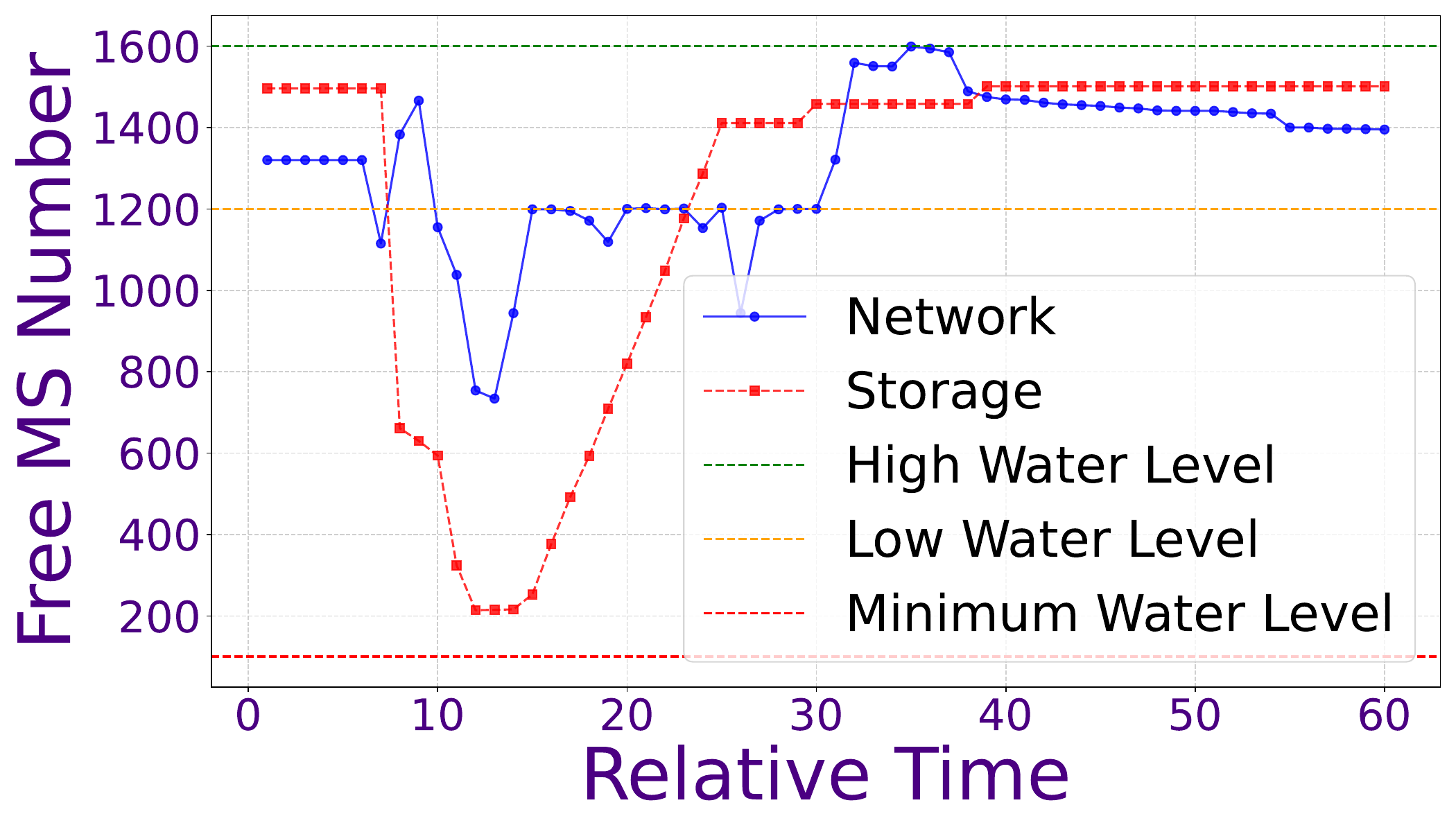}
        \subcaption{Differences in Waterlevel}
        \label{fig:waterlevel_upgrade}
    \end{minipage}\hfill
    \begin{minipage}{0.23\textwidth}
        \centering
        \includegraphics[width=\linewidth,height=3cm]{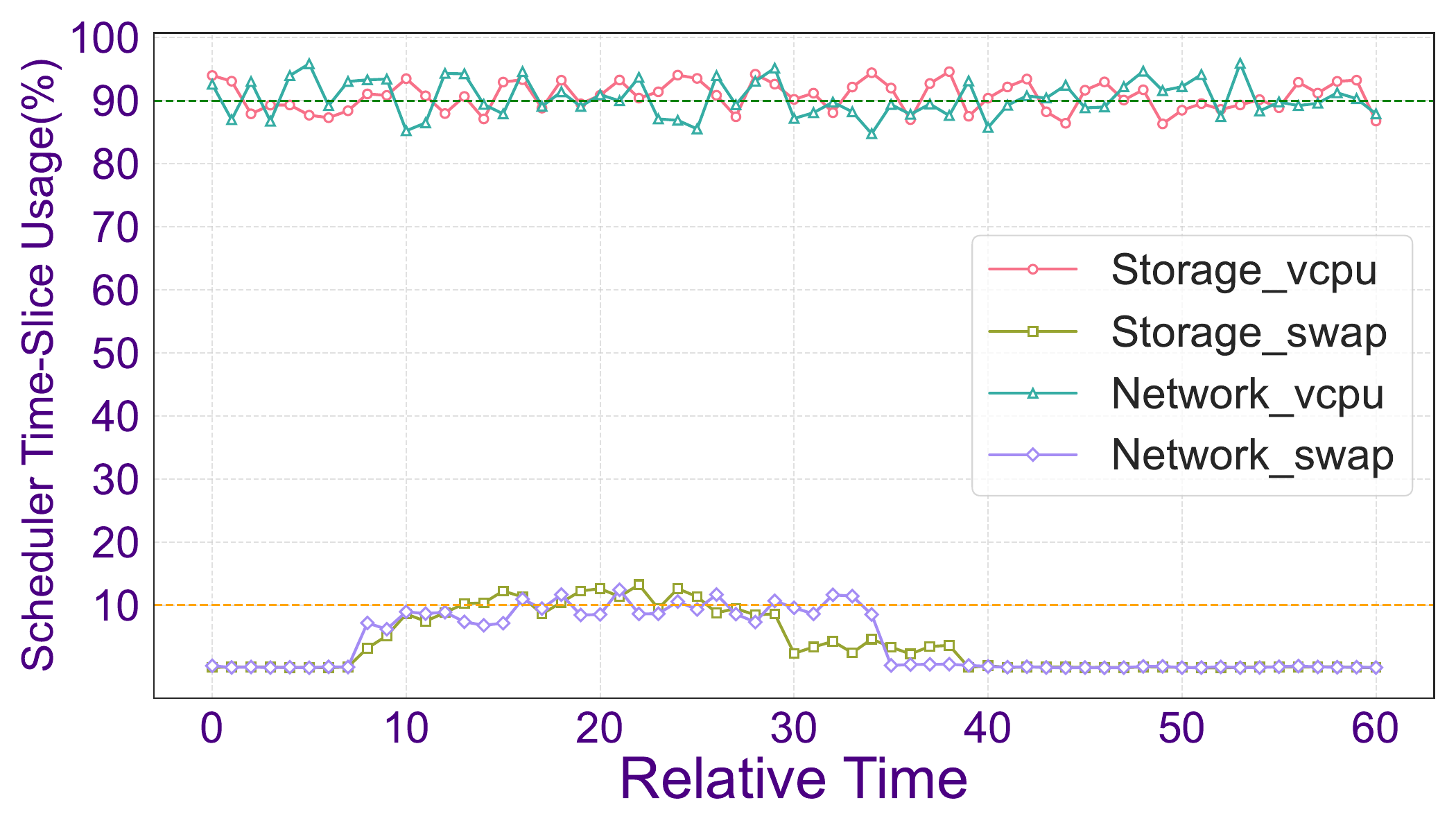}
        \\
        \subcaption{Scheduler Slices Ratio}
        \label{fig:sched_eval}
    \end{minipage}
    \begin{minipage}{0.23\textwidth}
        \centering
        \includegraphics[width=\linewidth,height=3cm]{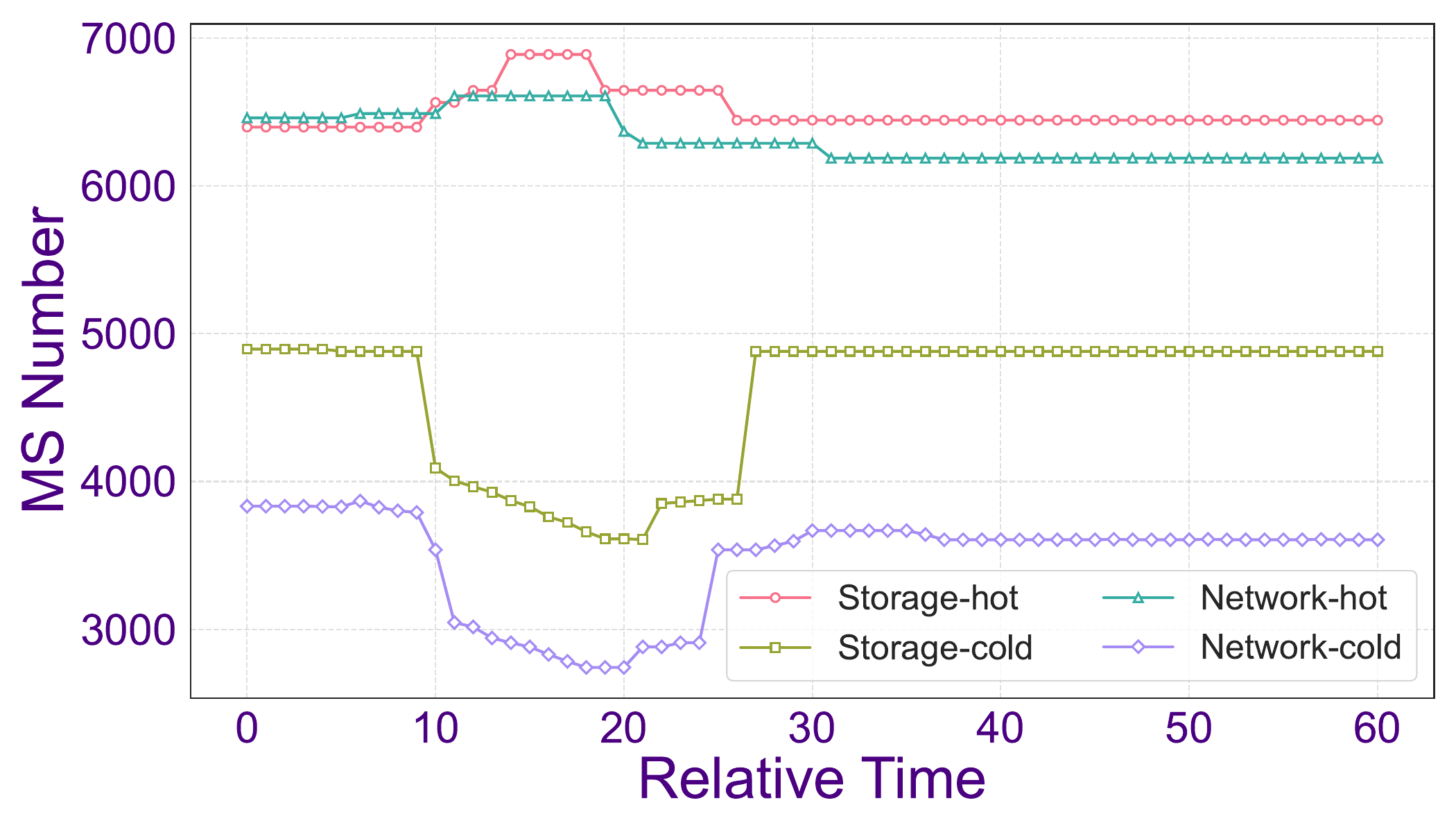}
        \subcaption{Hot and Cold MS Number.}
        \label{fig:lru_ms}
    \end{minipage}\hfill
    \begin{minipage}{0.23\textwidth}
        \centering
        \includegraphics[width=\linewidth,height=3cm]{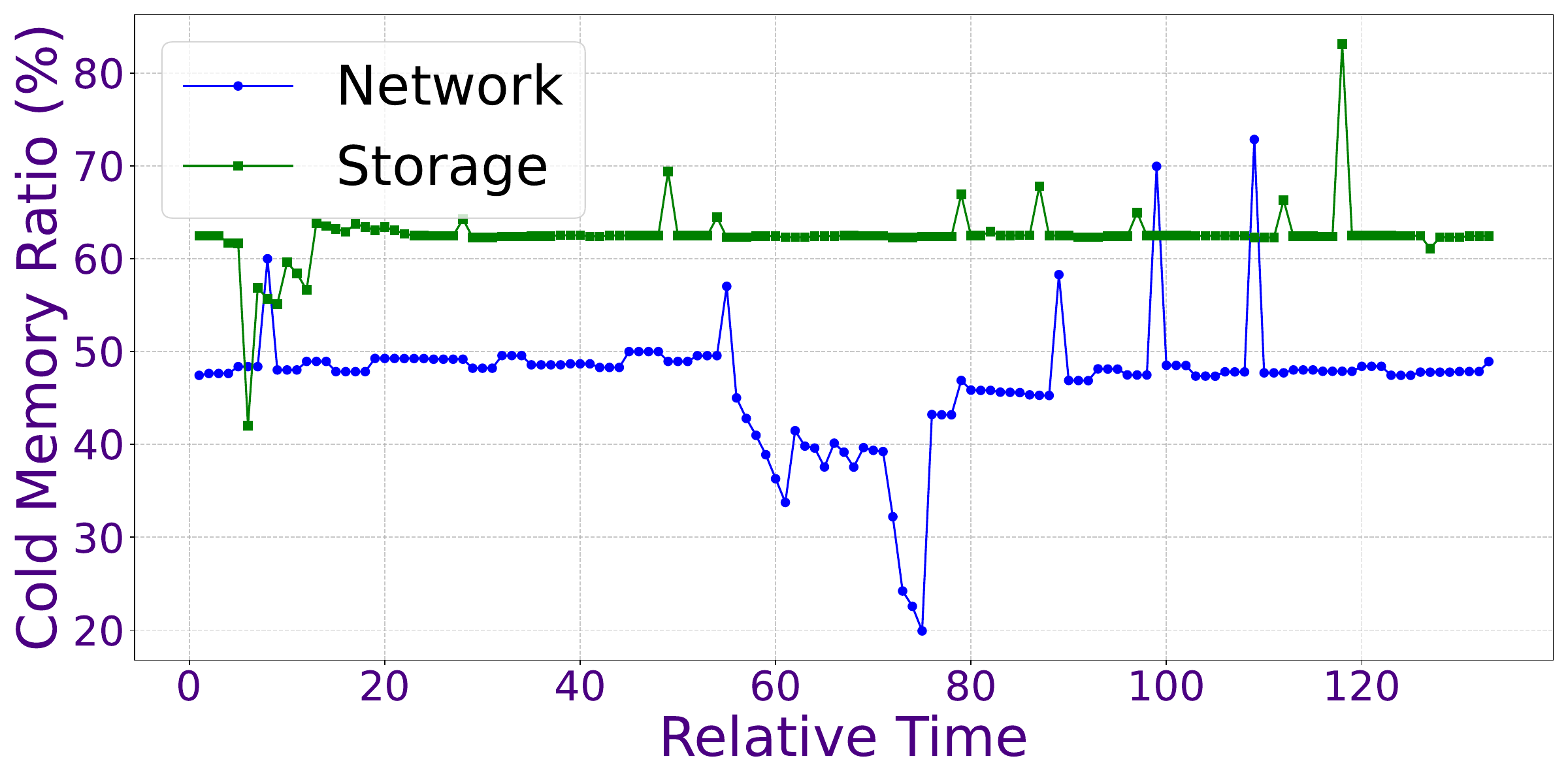}
        \\
        \subcaption{Cold Memory Ratio}
        \label{fig:lru_upgrade}
    \end{minipage}
    \begin{minipage}{0.23\textwidth}
        \centering
        \includegraphics[width=\linewidth,height=3cm]{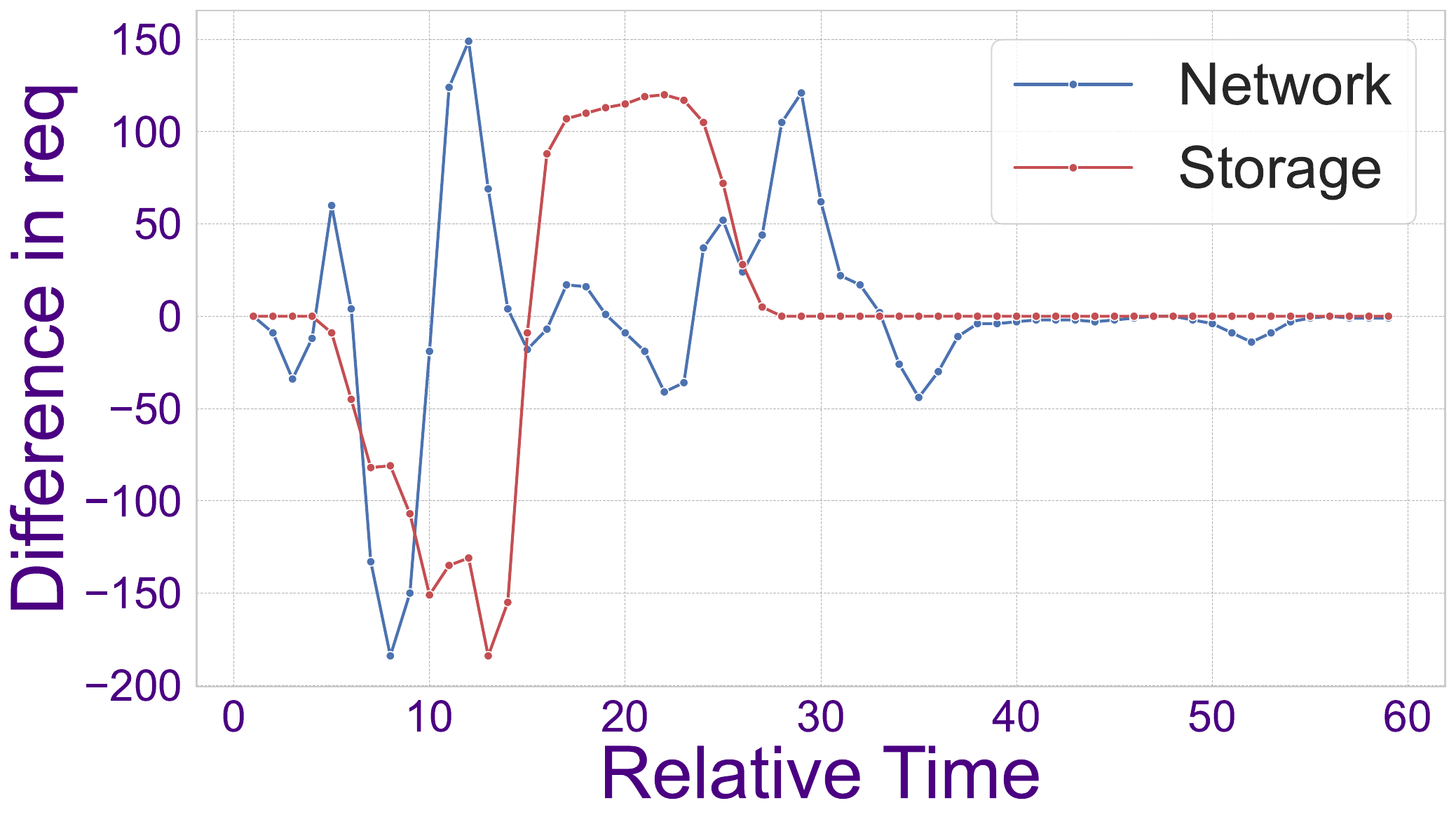}
        \subcaption{Differences in Swap \textit{req}}
        \label{fig:req_ms_upgrade}
    \end{minipage}\hfill
    \begin{minipage}{0.23\textwidth}
        \centering
        \includegraphics[width=\linewidth,height=3cm]{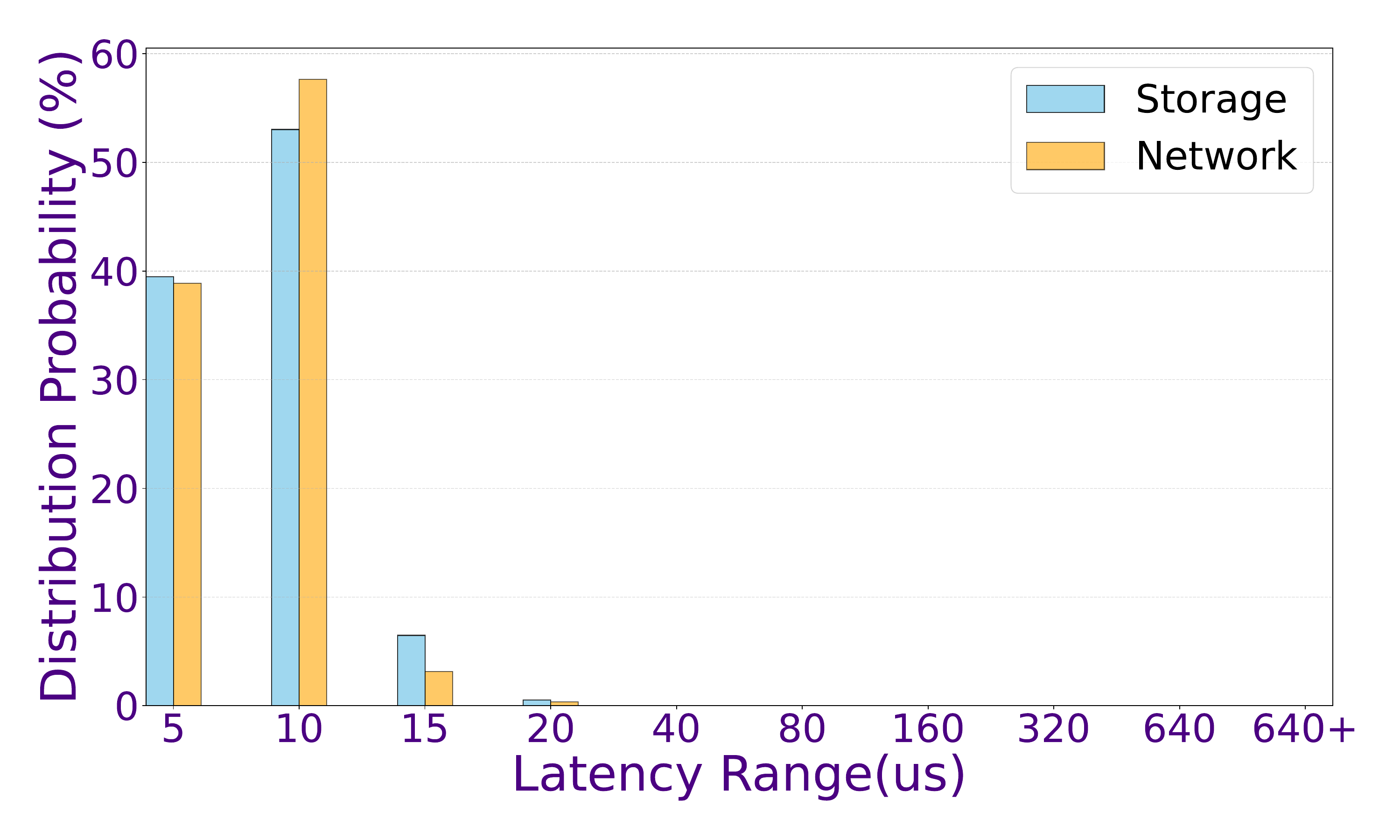}
        \\
        \subcaption{Latency of EPT Fault}
        \label{fig:ept_fault_upgrade}
    \end{minipage}
    \caption{The Running Cost and Benefit of Taiji.}
    \label{fig:overall}
\end{figure}

For high-load scenarios, we test hot upgrades for network and storage, where processes transfer up to 8\,GB of data and perform data copying and conversion between old and new processes. As shown in Figure~\ref{fig:req_ms_upgrade}, hot upgrades trigger memory access spikes, dropping free memory below the \texttt{low} threshold. Taiji proactively swaps until free memory exceeds \texttt{high}, adapting to workload changes. Even under high load, memory stays above \texttt{min}, preventing exhaustion and ensuring stability. As shown in Figure~\ref{fig:sched_eval}, during hot upgrade, fair scheduling ensures stable frontend vCPU execution with mostly consistent time-slice allocation. The backend swap task remains idle before upgrade due to meeting watermark thresholds, becomes active during upgrade, and is later constrained to the expected scheduling ratio. These results show Taiji’s front-end and back-end scheduling promptly dispatches elastic memory tasks to meet burst demands, achieving \textbf{O3}.

As shown in Figure~\ref{fig:lru_ms}, the hot page count remains stable during hot upgrade due to multi-level hot/cold tracking that filters transient accesses and prevents abrupt fluctuations. Increased memory access triggers proactive reclamation, reducing cold page count during the upgrade. Figure~\ref{fig:lru_upgrade} also shows cold data changes: it drops sharply as memory is swapped in and added to the \texttt{hot} set before stabilizing, with minor upward spikes from LRU cache refreshes. Storage stabilizes faster than network due to lower memory use, causing fewer faults and swaps. Even under intensive access, substantial reserved memory remains: for network, the cold-page ratio briefly drops to 20\% but stays above 35\%; for storage, it remains above 40\%. These results show that even under high load, cold pages can sustain DPU memory elasticity.

Figure~\ref{fig:req_ms_upgrade} shows \texttt{req} variation during hot upgrades as the difference from pre-upgrade values. Fluctuations are generally small. In network upgrades, an initial memory access spike causes heavy swap-in, lowering \texttt{req}; proactive swap-out then raises it, with smaller oscillations reflecting multiple intensive access phases, all within acceptable limits. In storage, \texttt{req} also drops initially due to large transfer memory use, triggering the water-level policy and proactive swap-out, after which \texttt{req} rises and stabilizes once demand is met. Unlike network, storage triggers the policy only once, indicating more concentrated access.

Figure~\ref{fig:ept_fault_upgrade} shows the EPT fault latency distribution during hot upgrades. Under large-scale memory access, over 90\% of EPT page fault latencies are below 10\,$\mu$s, with 92.51\% for storage and 95.50\% for network. No significant long-tail latency is observed. This shows that even under extreme load, Taiji sustains low swap-in latency, achieving \textbf{O2}.

\subsubsection{Online Operation Analysis}
\label{swapeval}
\begin{figure}[htbp]
    \centering
    \begin{minipage}{0.23\textwidth}
        \centering
        \includegraphics[width=\linewidth,height=3cm]{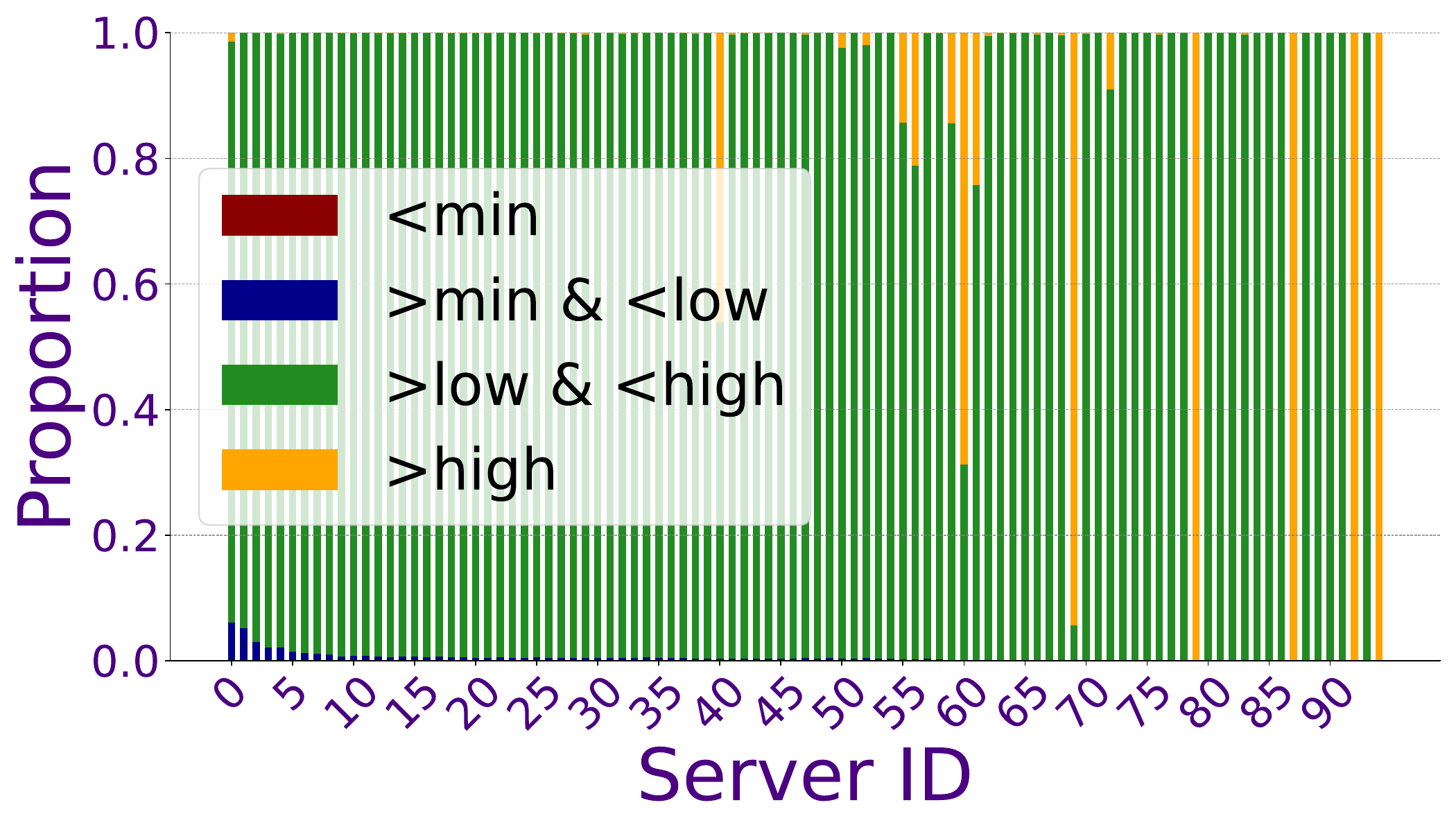}
        \subcaption{Waterlevel Distribution}
        \label{fig:waterlevel_cluster}
    \end{minipage}\hfill
    \begin{minipage}{0.23\textwidth}
        \centering
        \includegraphics[width=\linewidth,height=3cm]{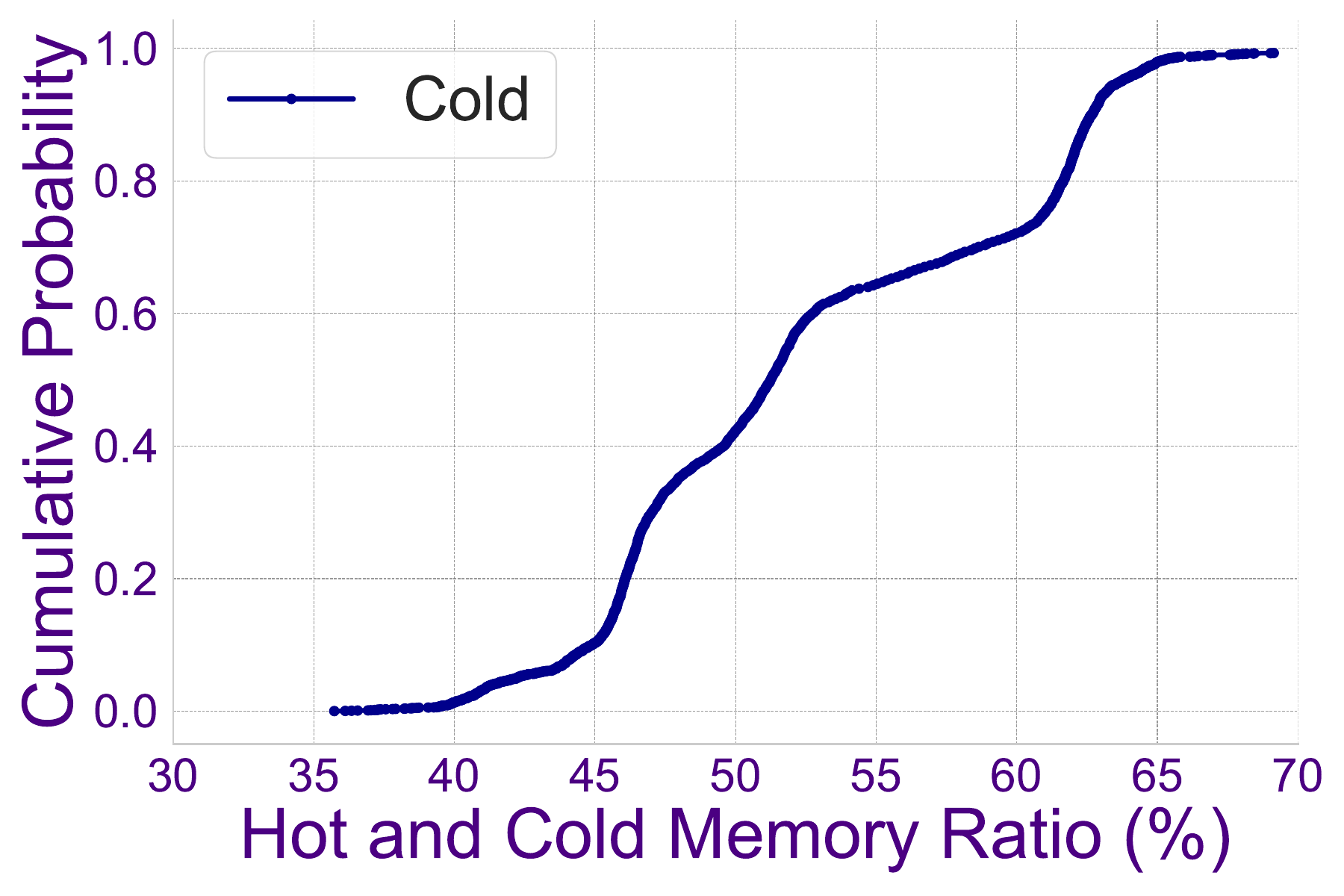}
        \\
        \subcaption{Cold Memory Ratio}
        \label{fig:lru_cluster}
    \end{minipage}
    \begin{minipage}{0.23\textwidth}
        \centering
        \includegraphics[width=\linewidth,height=3cm]{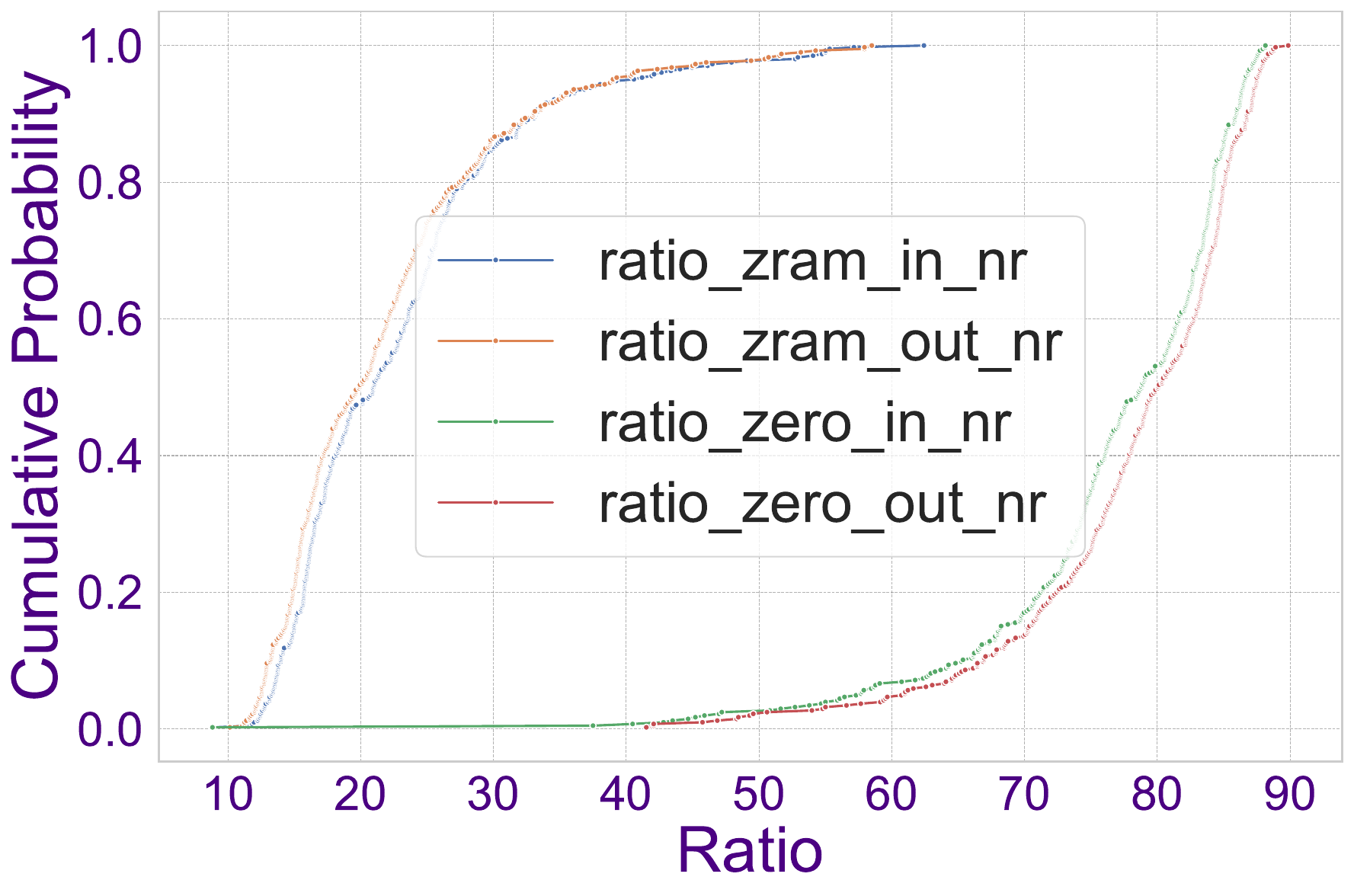}
        \subcaption{Swap-in/Swap-out Ratio}
        \label{fig:zram_zero_cluster}
    \end{minipage}\hfill
    \begin{minipage}{0.23\textwidth}
        \centering
        \includegraphics[width=\linewidth,height=3cm]{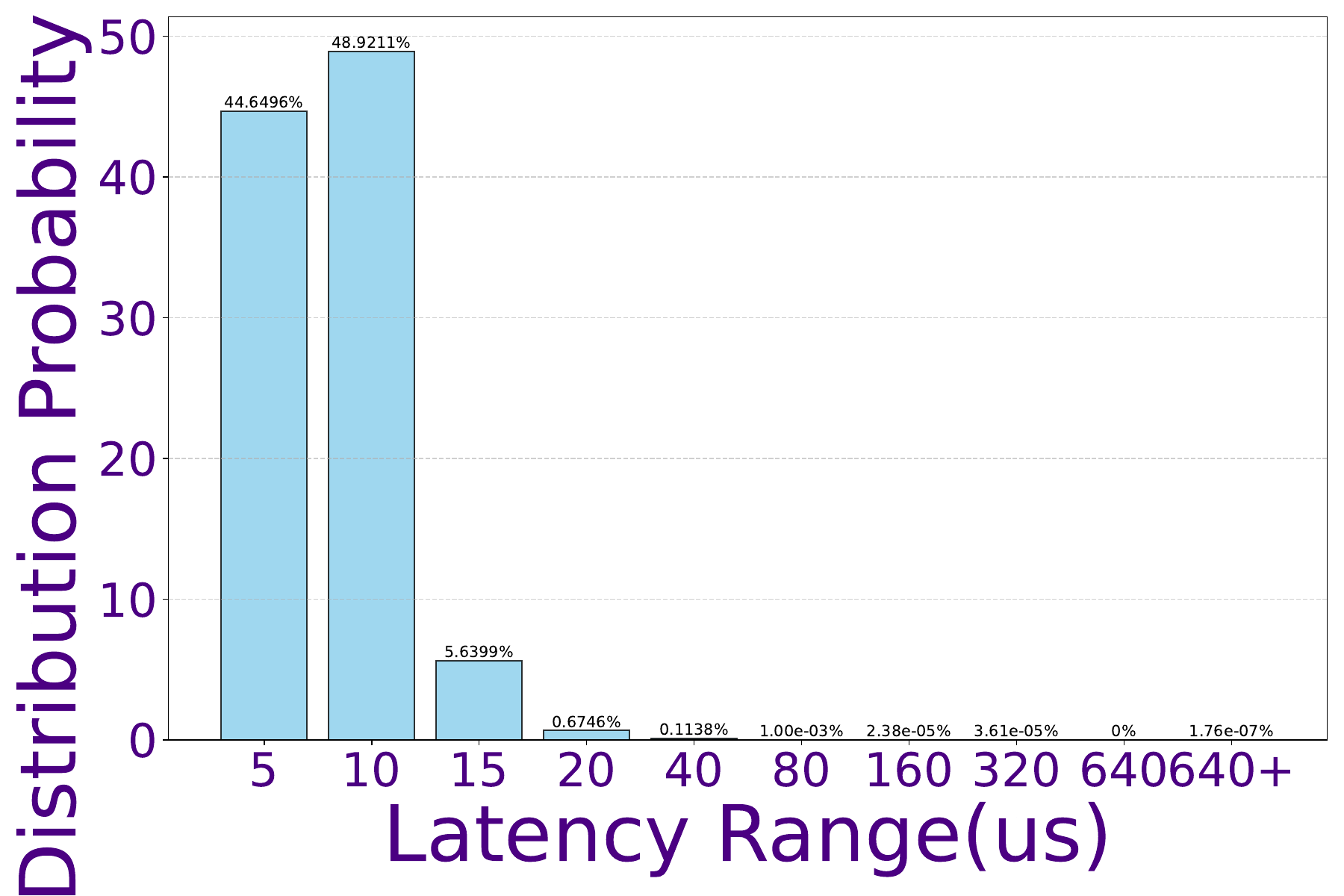}
        \\
        \subcaption{Latency of EPT Fault}
        \label{fig:ept_fault_cluster}
    \end{minipage}
    \caption{The Running Cost and Benefit of Taiji.}
    \label{fig:overall}
\end{figure}

To evaluate Taiji’s practical behavior, we analyze runtime memory swapping using real online data. Each DPU has 32\,GB physical and 16\,GB virtual memory, achieving 50\% elastic expansion. Figure~\ref{fig:waterlevel_cluster} shows memory water levels across cluster nodes over one hour: most remain between \texttt{low} and \texttt{high}, some above \texttt{high}, and only a few occasionally below \texttt{low} due to memory-intensive workloads like hot upgrades or traffic spikes. This indicates most nodes maintain safe levels, ensuring burst capacity and demonstrating Taiji’s water-level policy sustains diverse production workloads.

Figure~\ref{fig:lru_cluster} shows the CMF of the cold/hot memory ratio, indicating Taiji’s multi-level LRU accurately identifies cold pages. The cluster’s average cold-memory ratio is 52.79\%, and even the most memory-utilized nodes stay above 30\%, ensuring ample swap space for DPU expansion. Figure~\ref{fig:zram_zero_cluster} shows that of all swapped MPs, 76.79\% are zero pages and 23.21\% are compressed pages, with an average compression ratio of 47.63\%, greatly reducing their footprint. These results show Taiji’s elastic back-end memory reduces space overhead and improves overall memory utilization.

Figure~\ref{fig:ept_fault_cluster} shows the percentile distribution of EPT latencies across nodes during this period: 99\% are below 15\,$\mu$s, and 93.57\% are within 10\,$\mu$s, meeting \textbf{O2}. This confirms that Taiji’s page-fault latency distribution across the cluster meets design expectations, enabling timely swap-in to minimize application performance impact. Occasional higher latencies indicate room for improvement in EPT fault handling, which future optimizations should address.

\subsubsection{Overcommitting Benefit}
\label{benefit}

Taiji supports multiple elastic memory configurations, with the common case being a 32\,GB DPU with 16\,GB virtual elastic memory, achieving 50\% elasticity. Figure~\ref{fig:req_cluster} shows the CMF of swapped pages: about 50\% of nodes swap fewer than 8{,}000 MSes, and most fall between 6{,}000 and 11{,}000. Swapping 8{,}000 MSes frees 15.6\,GB of memory. The \texttt{Req} metadata overhead for recording swap information is about 20\,MB, including 15\,MB for CRC values to ensure correctness. From the compression analysis in Section~\ref{swapeval}, storing these swapped pages uses only 1.73\,GB, yielding a 9$\times$ overselling gain. Compared to the average metadata cost of 127.33\,MB or a reserved 400\,MB, the benefit-to-cost ratio reaches 125.5$\times$ or 39$\times$, fully meeting \textbf{O3}.

\section{Related Work}
With rising demands in cloud computing, big data, and AI, companies like AWS \cite{AWSNitro2024}, NVIDIA \cite{HPCRL2025,BlueFieldDPU_Blog}, and Intel introduced DPUs~\cite{RN26} to meet data center network/storage needs~\cite{RN26,RN23,RN7}. NVIDIA’s BlueField integrates ARM cores, programmable accelerators, and high-speed interfaces, offloading IO from CPUs to boost performance~\cite{RN7}. Programmability supports diverse workloads; in clouds, DPUs reduce host CPU load and improve efficiency~\cite{RN2}. Yet DPU performance faces challenges: complex hardware and slow upgrades limit adaptability~\cite{RN25,RN26}. Growing memory/CPU demands and slow upgrade cycles make elasticity urgent. Existing techniques target cloud services but neglect DPU-specific needs~\cite{RN37,RN40}. With many DPUs deployed, hot-swappable elasticity is needed.

Memory overcommitment allocates virtual machines more memory than the host’s physical capacity to optimize resource use~\cite{chang2024gmt}. If mismanaged, it can degrade performance~\cite{RN12}, and reclamation techniques such as ballooning cause slowdowns from increased paging activity and workload variability~\cite{RN15}. While ESXi\cite{ESXi}, KVM, and Xen support overcommitment, vendors discourage its use to avoid unpredictable performance loss~\cite{RN8,RN9}. Double paging by both hypervisor and guest OS adds severe overhead, and in containerized environments, overcommitment also significantly impacts performance~\cite{RN12}. Substantial improvements are required for elastic memory to be practical at production scale.

Researchers proposed methods for memory overcommitment: MemFlex~\cite{zhang2017memflex} enhances guest swapping; Zweilous~\cite{Zweilous} improves compression; PMA~\cite{PMA} reclaims inactive memory with <$10\%$ loss and 33\% overcommitment; HeMem~\cite{HeMem} manages tiered memory asynchronously; Pond~\cite{Pond} builds CXL pools with ML-based prediction. Application-aware approaches~\cite{RN18,RN17} also maintain acceptable performance. TMO~\cite{weiner2022tmo} measures work loss and adjusts offloading in real time. However, in high-performance DPUs, overcommitment still suffers from unpredictable degradation and complex management~\cite{PMA}. Thus, lightweight, scalable, high-performance virtualization for DPUs is urgently needed.

\section{Discussion}
\subsection{Data Correctness}
A key factor in DPU performance is extensive DMA usage, but as current DMA devices lack retry support, swapping memory must be avoided to prevent corruption. Binding all device-used I/O memory leaves too little movable memory to expand virtual memory by over 50\% (\textbf{O3}). Since this DMA-related memory is dynamic, Taiji lets applications specify DMA ranges for protection and ensures timely swap-in before access. Taiji also intercepts DMAR exceptions and uses CRC to ensure correctness. Future hardware–software co-design will enable DMA-triggered swap-ins without application tags. These mechanisms allow Taiji to match \textbf{O2}’s efficiency without compromising data correctness.

\subsection{Backend Memory}
Taiji’s backend storage supports zero, compressed, free pages, remote memory, and disks. In the guest OS, it detects swapped free pages and reclaims their page tables immediately, with later swap-in only rebuilding tables—enhancing efficiency and aiding \textbf{O2}’s low-latency goal. As such pages are rare and detection incurs zone lock overhead that disrupts the guest OS, this type is disabled in production. Remote memory and disks act as fallback storage for bursts beyond Taiji’s elasticity, though none have occurred; the common case is hot upgrades temporarily doubling memory, which current swapping handles. These backends remain optional.

\subsection{Portability}
Taiji is currently implemented on DPUs with Intel’s x86 CPU, whose virtualization is closely tied to the underlying hardware, while elastic memory is hardware-agnostic. By adapting hardware-specific features and reusing other components, Taiji can be ported to other virtualization platforms such as AMD and ARM. For AMD, adaptations include differences between SVM and VMX, IOMMU formats, LAPIC, and timer control. Once addressed, CPU virtualization can run smoothly. Porting to ARM is similar but requires additional work due to substantial differences in registers and virtualization modes from x86. Resolving these hardware differences enables Taiji to be ported to other DPU platforms.

\subsection{Extensibility}
In scheduling, CPs yield capacity to elastic tasks, but under high load or growing demands, resource shortages may occur. Taiji’s extensible scheduler can create VCPUs on demand by initializing a VMCS without register passthrough. When hot-plugging a VCPU, the virtualization layer detects its startup signal and, with the scheduler allocating time slices, schedules the VCPU task. The new VCPU behaves like a switched VCPU and, with proper time slices, runs normally in the virtualization layer, transparently adding CPU resources for elastic scaling. Coordinating and fully utilizing physical CPUs via the scheduler, Taiji unifies memory and CPU elasticity while reducing maintenance cost.

\section{Conclusion and Future Work}
We present Taiji, a lightweight DPU elasticity solution that inserts a virtualization layer between OS and hardware, enabling full memory swapping and elastic scaling. Leveraging multi-level LRUs, parallel swapping, and transparent scheduling, Taiji achieves efficient memory elasticity. Its modular design supports hot switching and upgrades, improving availability on existing and new DPUs. Experiments show Taiji improves utilization, lowers hardware cost, and raises service capacity while sustaining high performance. Future work includes porting beyond x86, enhancing CPU elasticity, and eliminating user-space DMA tagging.


\bibliographystyle{ACM-Reference-Format}
\bibliography{taiji}

\end{document}